\pdfoutput=1

\documentclass[12pt,a4paper]{article}
\usepackage{ifthen} 
\newboolean{pdflatex}
\setboolean{pdflatex}{true} 

\usepackage{afterpage}

\newboolean{articletitles}
\setboolean{articletitles}{true} 

\newboolean{uprightparticles}
\setboolean{uprightparticles}{false} 

\newboolean{inbibliography}
\setboolean{inbibliography}{false} 

\usepackage[top=1in, bottom=1.25in, left=1in, right=1in]{geometry}

\usepackage{microtype}
\usepackage{lineno}  
\usepackage{xspace} 
\usepackage{caption}

\usepackage{graphicx}  
\usepackage{color}
\usepackage{colortbl}
\graphicspath{{./figs/}} 

\usepackage{amsmath} 
\usepackage{amssymb}
\usepackage{amsfonts}
\usepackage{upgreek} 

\newcommand*\patchAmsMathEnvironmentForLineno[1]{%
\expandafter\let\csname old#1\expandafter\endcsname\csname #1\endcsname
\expandafter\let\csname oldend#1\expandafter\endcsname\csname
end#1\endcsname
 \renewenvironment{#1}%
   {\linenomath\csname old#1\endcsname}%
   {\csname oldend#1\endcsname\endlinenomath}%
}
\newcommand*\patchBothAmsMathEnvironmentsForLineno[1]{%
  \patchAmsMathEnvironmentForLineno{#1}%
  \patchAmsMathEnvironmentForLineno{#1*}%
}
\AtBeginDocument{%
\patchBothAmsMathEnvironmentsForLineno{equation}%
\patchBothAmsMathEnvironmentsForLineno{align}%
\patchBothAmsMathEnvironmentsForLineno{flalign}%
\patchBothAmsMathEnvironmentsForLineno{alignat}%
\patchBothAmsMathEnvironmentsForLineno{gather}%
\patchBothAmsMathEnvironmentsForLineno{multline}%
\patchBothAmsMathEnvironmentsForLineno{eqnarray}%
}

\usepackage{hyperref}    
\usepackage[all]{hypcap} 
\usepackage{xspace} 
\usepackage{upgreek}

\def\lhcb {\mbox{LHCb}\xspace}

\def\belle  {\mbox{Belle}\xspace}

\def\MagUp {\mbox{\em Mag\kern -0.05em Up}\xspace}

\ifthenelse{\boolean{uprightparticles}}%
{

 \def\Pmu         {\ensuremath{\upmu}\xspace}

 \def\Ppi         {\ensuremath{\uppi}\xspace}

 \def\Ppsi        {\ensuremath{\uppsi}\xspace}

 \def\PDelta      {\ensuremath{\Delta}\xspace}                 
 \def\PXi      {\ensuremath{\Xi}\xspace}                 
 \def\PLambda      {\ensuremath{\Lambda}\xspace}                 
 \def\PSigma      {\ensuremath{\Sigma}\xspace}                 
 \def\POmega      {\ensuremath{\Omega}\xspace}                 
 \def\PUpsilon      {\ensuremath{\Upsilon}\xspace}                 
                  
 \def\PB      {\ensuremath{\mathrm{B}}\xspace}                 
                  
 \def\PD      {\ensuremath{\mathrm{D}}\xspace}

 \def\PJ      {\ensuremath{\mathrm{J}}\xspace}                 
 \def\PK      {\ensuremath{\mathrm{K}}\xspace}

 \def\Pb      {\ensuremath{\mathrm{b}}\xspace}                 
 \def\Pc      {\ensuremath{\mathrm{c}}\xspace}

 \def\Pi      {\ensuremath{\mathrm{i}}\xspace}

 \def\Pp      {\ensuremath{\mathrm{p}}\xspace}

 \def\Ps      {\ensuremath{\mathrm{s}}\xspace}

}
{

 \def\Pmu         {\ensuremath{\mu}\xspace}

 \def\Ppi         {\ensuremath{\pi}\xspace}

 \def\Ppsi        {\ensuremath{\psi}\xspace}                 
                  
 \mathchardef\PDelta="7101
 \mathchardef\PXi="7104
 \mathchardef\PLambda="7103
 \mathchardef\PSigma="7106
 \mathchardef\POmega="710A
 \mathchardef\PUpsilon="7107
                  
 \def\PB      {\ensuremath{B}\xspace}                 
                  
 \def\PD      {\ensuremath{D}\xspace}

 \def\PJ      {\ensuremath{J}\xspace}                 
 \def\PK      {\ensuremath{K}\xspace}

 \def\Pb      {\ensuremath{b}\xspace}                 
 \def\Pc      {\ensuremath{c}\xspace}

 \def\Pi      {\ensuremath{i}\xspace}

 \def\Pp      {\ensuremath{p}\xspace}

 \def\Ps      {\ensuremath{s}\xspace}

}

\makeatletter
\ifcase \@ptsize \relax
  \newcommand{\miniscule}{\@setfontsize\miniscule{4}{5}}
\or
  \newcommand{\miniscule}{\@setfontsize\miniscule{5}{6}}
\or
  \newcommand{\miniscule}{\@setfontsize\miniscule{5}{6}}
\fi
\makeatother

\DeclareRobustCommand{\optbar}[1]{\shortstack{{\miniscule (\rule[.5ex]{1.25em}{.18mm})}
  \\ [-.7ex] $#1$}}

\def\mumu       {{\ensuremath{\Pmu^+\Pmu^-}}\xspace}

\def\squark    {{\ensuremath{\Ps}}\xspace}

\def\cquark    {{\ensuremath{\Pc}}\xspace}

\def\bquark    {{\ensuremath{\Pb}}\xspace}
\def\bquarkbar {{\ensuremath{\overline \bquark}}\xspace}

\def\pion   {{\ensuremath{\Ppi}}\xspace}

\def\pip    {{\ensuremath{\pion^+}}\xspace}
\def\pim    {{\ensuremath{\pion^-}}\xspace}

\def\kaon    {{\ensuremath{\PK}}\xspace}
  \def\Kbar    {{\kern 0.2em\overline{\kern -0.2em \PK}{}}\xspace}

\def\KorKbar    {\kern 0.18em\optbar{\kern -0.18em K}{}\xspace}

\def\Kp      {{\ensuremath{\kaon^+}}\xspace}
\def\Km      {{\ensuremath{\kaon^-}}\xspace}

\def\Kstarzb {{\ensuremath{\Kbar{}^{*0}}}\xspace}

  \def\Dbar    {{\kern 0.2em\overline{\kern -0.2em \PD}{}}\xspace}
\def\D       {{\ensuremath{\PD}}\xspace}

\def\DorDbar    {\kern 0.18em\optbar{\kern -0.18em D}{}\xspace}

\def\Dsp     {{\ensuremath{\D^+_\squark}}\xspace}
\def\Dsm     {{\ensuremath{\D^-_\squark}}\xspace}

\def\B       {{\ensuremath{\PB}}\xspace}
\def\Bbar    {{\ensuremath{\kern 0.18em\overline{\kern -0.18em \PB}{}}}\xspace}

\def\BorBbar    {\kern 0.18em\optbar{\kern -0.18em B}{}\xspace}
\def\Bz      {{\ensuremath{\B^0}}\xspace}
\def\Bzb     {{\ensuremath{\Bbar{}^0}}\xspace}
\def\Bu      {{\ensuremath{\B^+}}\xspace}

\def\Bd      {{\ensuremath{\B^0}}\xspace}
\def\Bs      {{\ensuremath{\B^0_\squark}}\xspace}
\def\Bsb     {{\ensuremath{\Bbar{}^0_\squark}}\xspace}
\def\Bdb     {{\ensuremath{\Bbar{}^0}}\xspace}
\def\Bc      {{\ensuremath{\B_\cquark^+}}\xspace}

\def\jpsi     {{\ensuremath{{\PJ\mskip -3mu/\mskip -2mu\Ppsi\mskip 2mu}}}\xspace}

 \def\Y#1S{\ensuremath{\PUpsilon{(#1S)}}\xspace}

\def\proton      {{\ensuremath{\Pp}}\xspace}

\def\Lz          {{\ensuremath{\PLambda}}\xspace}
\def\Lbar        {{\ensuremath{\kern 0.1em\overline{\kern -0.1em\PLambda}}}\xspace}
\def\LorLbar    {\kern 0.18em\optbar{\kern -0.18em \PLambda}{}\xspace}

\def\Lb      {{\ensuremath{\Lz^0_\bquark}}\xspace}

\def\to                 {\ensuremath{\rightarrow}\xspace}

\def\CP                {{\ensuremath{C\!P}}\xspace}

\newcommand{\dms}{{\ensuremath{\Delta m_{\squark}}}\xspace}

\newcommand{\DGs}{{\ensuremath{\Delta\Gamma_{\squark}}}\xspace}

\newcommand{\Gs}{{\ensuremath{\Gamma_{\squark}}}\xspace}

\newcommand{\GL}{{\ensuremath{\Gamma_{\mathrm{ L}}}}\xspace}
\newcommand{\GH}{{\ensuremath{\Gamma_{\mathrm{ H}}}}\xspace}

\newcommand{\phis}{{\ensuremath{\phi_{\squark}}}\xspace}

\def\AT#1     {\ensuremath{A_{\mathrm{T}}^{#1}}\xspace}

\def\C#1      {\ensuremath{\mathcal{C}_{#1}}\xspace}                       
\def\Cp#1     {\ensuremath{\mathcal{C}_{#1}^{'}}\xspace}                    
\def\Ceff#1   {\ensuremath{\mathcal{C}_{#1}^{\mathrm{(eff)}}}\xspace}        
\def\Cpeff#1  {\ensuremath{\mathcal{C}_{#1}^{'\mathrm{(eff)}}}\xspace}       
\def\Ope#1    {\ensuremath{\mathcal{O}_{#1}}\xspace}                       
\def\Opep#1   {\ensuremath{\mathcal{O}_{#1}^{'}}\xspace}

\newcommand{\tev}{\ifthenelse{\boolean{inbibliography}}{\ensuremath{~T\kern -0.05em eV}}{\ensuremath{\mathrm{\,Te\kern -0.1em V}}}\xspace}
\newcommand{\gev}{\ensuremath{\mathrm{\,Ge\kern -0.1em V}}\xspace}
\newcommand{\mev}{\ensuremath{\mathrm{\,Me\kern -0.1em V}}\xspace}
\newcommand{\kev}{\ensuremath{\mathrm{\,ke\kern -0.1em V}}\xspace}
\newcommand{\ev}{\ensuremath{\mathrm{\,e\kern -0.1em V}}\xspace}
\newcommand{\gevc}{\ensuremath{{\mathrm{\,Ge\kern -0.1em V\!/}c}}\xspace}
\newcommand{\mevc}{\ensuremath{{\mathrm{\,Me\kern -0.1em V\!/}c}}\xspace}
\newcommand{\gevcc}{\ensuremath{{\mathrm{\,Ge\kern -0.1em V\!/}c^2}}\xspace}
\newcommand{\gevgevcccc}{\ensuremath{{\mathrm{\,Ge\kern -0.1em V^2\!/}c^4}}\xspace}
\newcommand{\mevcc}{\ensuremath{{\mathrm{\,Me\kern -0.1em V\!/}c^2}}\xspace}

\def\m    {\ensuremath{\mathrm{ \,m}}\xspace}

\def\mum  {\ensuremath{{\,\upmu\mathrm{m}}}\xspace}

\def\invfb   {\ensuremath{\mbox{\,fb}^{-1}}\xspace}

\def\ps   {\ensuremath{{\mathrm{ \,ps}}}\xspace}
\def\fs   {\ensuremath{\mathrm{ \,fs}}\xspace}

\def\invps{\ensuremath{{\mathrm{ \,ps^{-1}}}}\xspace}

\newcommand{\stat}{\ensuremath{\mathrm{\,(stat)}}\xspace}

\newcommand{\chisq}{\ensuremath{\chi^2}\xspace}

\newcommand{\chisqip}{\ensuremath{\chi^2_{\text{IP}}}\xspace}

\def\gsim{{~\raise.15em\hbox{$>$}\kern-.85em
          \lower.35em\hbox{$\sim$}~}\xspace}
\def\lsim{{~\raise.15em\hbox{$<$}\kern-.85em
          \lower.35em\hbox{$\sim$}~}\xspace}

\newcommand{\mean}[1]{\ensuremath{\left\langle #1 \right\rangle}} 
\newcommand{\Real}{\ensuremath{\mathcal{R}e}\xspace}
\newcommand{\Imag}{\ensuremath{\mathcal{I}m}\xspace}

\def\sPlot{\mbox{\em sPlot}\xspace}

\def\ptot       {\mbox{$p$}\xspace}
\def\pt         {\mbox{$p_{\mathrm{ T}}$}\xspace}

\def\evtgen     {\mbox{\textsc{EvtGen}}\xspace}

\def\geant      {\mbox{\textsc{Geant4}}\xspace}

\def\photos     {\mbox{\textsc{Photos}}\xspace}

\def\pythia     {\mbox{\textsc{Pythia}}\xspace}

\def\tell1  {TELL1\xspace}
\def\ukl1   {UKL1\xspace}

\newcommand{\ie}{\mbox{\itshape i.e.}\xspace}

\usepackage{cite} 
\usepackage{mciteplus}

\newcommand{\mygevc}{\ensuremath{{\mathrm{Ge\kern -0.1em V\!/}c}}\xspace}
\newcommand{\xx}{\ensuremath{\kern 0.5em }}

\newcommand{\BzbJpsiKpi}{\ensuremath{\Bzb\to\jpsi\Km \pip}\xspace}
\newcommand{\LbJpsipK}{\ensuremath{\Lb\to\jpsi\proton\Km}\xspace}

\newcommand{\Bsjpsikk}{\ensuremath{\Bs \to\jpsi \Kp \Km}\xspace}
\newcommand{\BorBb}{\Bs}

\def\sPlot{\mbox{\em sPlot}\xspace}
\def\sWeight{\mbox{\em sWeight}\xspace}

\usepackage{multirow} 
\usepackage{booktabs} 
\usepackage{rotating}

\def \m {m_{KK}}
\def \angmu {\theta_{\jpsi}}
\def \angpi {\theta_{KK}}

\def \Bq {B_s^{0}}
\def \Bqb {\overline{B}{}_s^{0}}

\def \ch {\cosh \frac{\DGs t}{2}}
\def \sh {\sinh \frac{\DGs t}{2}}
\def \Ab {\overline{A}}
\def \cs {\cos(\dms t)}
\def \sn {\sin(\dms t)}

\def \A {{\cal A}}
\def \cAb {\overline{{\cal A}}}

\DeclareRobustCommand{\optbar}[1]{\shortstack{{\miniscule (\rule[.5ex]{1.25em}{.18mm})}
  \\ [-.7ex] $#1$}}

\usepackage{epsfig,accents}

\usepackage{longtable} 
\begin{document}
\renewcommand{\thefootnote}{\fnsymbol{footnote}}
\setcounter{footnote}{1}

\begin{titlepage}
\pagenumbering{roman}

\vspace*{-1.5cm}
\centerline{\large EUROPEAN ORGANIZATION FOR NUCLEAR RESEARCH (CERN)}
\vspace*{1.5cm}
\noindent
\begin{tabular*}{\linewidth}{lc@{\extracolsep{\fill}}r@{\extracolsep{0pt}}}
\ifthenelse{\boolean{pdflatex}}
{\vspace*{-2.7cm}\mbox{\!\!\!\includegraphics[width=.14\textwidth]{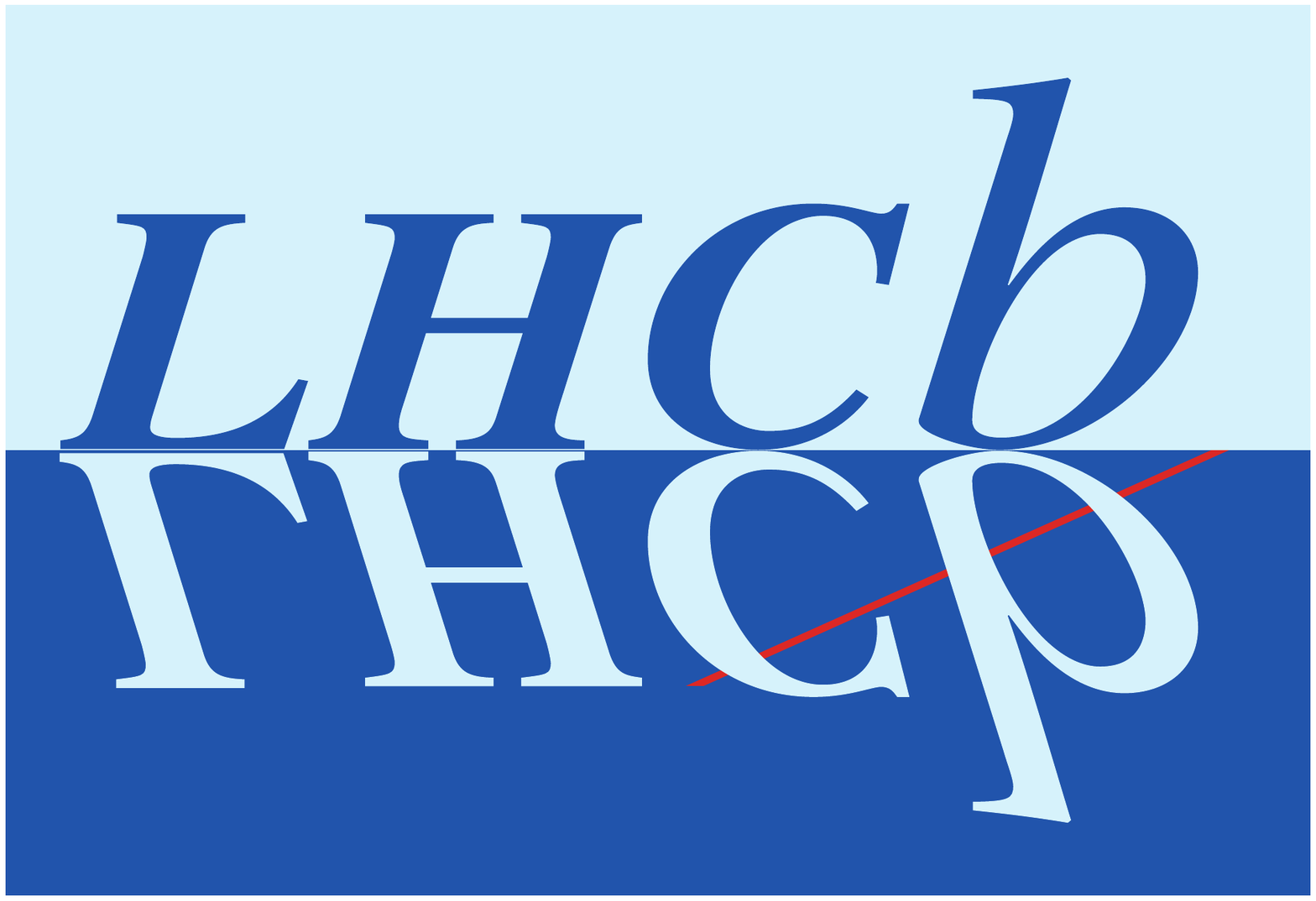}} & &}%
{\vspace*{-1.2cm}\mbox{\!\!\!\includegraphics[width=.12\textwidth]{lhcb-logo.eps}} & &}%
\\
 & & CERN-EP-2017-062 \\  
 & & LHCb-PAPER-2017-008 \\  
 & & April 26, 2017 \\ 
\end{tabular*}

\vspace*{4.0cm}

{\normalfont\bfseries\boldmath\huge
\begin{center}
Resonances and \CP violation in $\Bs$ and $\Bsb \to \jpsi K^+K^-$ decays in the mass region above the $\phi(1020)$
  \end{center}
}

\vspace*{2.0cm}

\begin{center}
The LHCb collaboration\footnote{Authors are listed at the end of this paper.}
\end{center}

\vspace{\fill}

\begin{abstract}
  \noindent
The decays of \Bs and \Bsb mesons into the $\jpsi K^+K^-$ final state are studied in the $K^+K^-$ mass region above the $\phi(1020)$ meson in order to determine the resonant substructure and measure the \CP-violating phase, $\phis$,  the decay width, $\Gs$, and the width difference between light and heavy mass eigenstates, $\DGs$.  A decay-time dependent amplitude analysis is employed. The data sample corresponds to an integrated luminosity of 3\invfb produced in 7 and 8\tev $pp$ collisions at the LHC, collected by the \lhcb experiment. The measurement determines $\phi_s = 119\pm107\pm34 {\rm \, mrad}$. A combination with previous LHCb measurements using similar decays into the  $\jpsi \pi^+\pi^-$ and $\jpsi\phi(1020)$ final states gives $\phis=1\pm37$\,mrad, consistent with the Standard Model prediction. 
  
\end{abstract}

\vspace*{2.0cm}

\begin{center}
Published in JHEP 08 (2017) 037
\end{center}

\vspace{\fill}

{\footnotesize 
\centerline{\copyright~CERN on behalf of the \lhcb collaboration, licence \href{http://creativecommons.org/licenses/by/4.0/}{CC-BY-4.0}.}}
\vspace*{2mm}

\end{titlepage}


\newpage
\setcounter{page}{2}
\mbox{~}
%
%
%
%

\cleardoublepage


\renewcommand{\thefootnote}{\arabic{footnote}}
\setcounter{footnote}{0}



\pagestyle{plain} 
\setcounter{page}{1}
\pagenumbering{arabic}


%


\section{Introduction}
\label{sec:Introduction}
Measurements of \CP violation through the interference of \Bs mixing and decay amplitudes are particularly sensitive to the presence of unseen particles or forces. The Standard Model (SM) prediction of the \CP-violating phase in quark-level $b\to c\overline{c}s$ transitions is very small, $\phi_s^{\rm SM}\equiv -2{\rm arg}\left(-\frac{V_{ts}V_{tb}^*}{V_{cs}V_{cb}^*}\right)\!=\!-36.5_{-1.2}^{+1.3}$\,mrad~\cite{Charles:2015gya}. 
Although subleading corrections from penguin amplitudes are ignored in 
this estimate, the interpretation of the current measurements is not affected, 
since those subleading terms are known to be small \cite{Fleischer:2015mla,LHCb-PAPER-2014-058,LHCB-PAPER-2015-034} compared to the experimental precision. Initial measurements 
 of $\phi_s$ were performed at the Tevatron \cite{Abazov:2011ry,*Aaltonen:2012ie}, followed by \lhcb measurements using both \Bs and \Bsb decays\footnote{Whenever a 
 flavour-specific decay is mentioned it also implies use of the charge-conjugate decay except when dealing with 
 \CP-violating quantities or other explicitly mentioned cases.} into $\jpsi\pi^+\pi^-$ and $\jpsi K^+K^-$, with $K^+K^-$ invariant masses\footnote{Natural units are used where $\hbar$=c=1.}  $\m<1.05$\gev,  from 3\invfb of integrated luminosity.  The measurements were found to be consistent with the SM value \cite{LHCb-PAPER-2014-059,LHCb-PAPER-2014-019}, as are more recent and somewhat less accurate results from the CMS~\cite{Khachatryan:2015nza} and ATLAS~\cite{Aad:2016tdj}  collaborations 
using $\jpsi\phi(1020)$ final states.\footnote{The final states $\Dsp\Dsm$~\cite{LHCb-PAPER-2014-051} and $\psi(2S)\phi(1020)$~\cite{LHCb-PAPER-2016-027} are also used by \lhcb, but the precisions are not comparable due to lower statistics.} The average of all of the above mentioned measurements is $\phis=-30\pm33$\,mrad~\cite{HFAG}.

Previously, using a data sample corresponding to 1\invfb integrated luminosity, the LHCb collaboration studied the resonant structures in the $\Bs\to\jpsi K^+K^-$ decay~\cite{LHCb-PAPER-2012-040} revealing a rich resonance spectrum in the $K^+K^-$ mass distribution. In addition to the $\phi(1020)$ meson, there are significant contributions  from the $f_2'(1525)$ resonance~\cite{LHCb-PAPER-2011-026} and nonresonant S-wave, which are large enough to allow further studies of \CP violation. This paper presents the first measurement of $\phi_s$ using $\Bs\to\jpsi K^+ K^-$ decays with  $\m$ above the $\phi(1020)$ region, using data corresponding to an integrated luminosity of 3\invfb, obtained from $pp$ collisions at the LHC. One third of the data was collected at a centre-of-mass energy of 7\tev, and the remainder at 8\tev. An amplitude analysis as a function of the $\Bs$ proper decay time~\cite{Zhang:2012zk} is performed to determine the \CP-violating phase $\phis$, by measuring simultaneously the \CP-even and \CP-odd decay amplitudes for each contributing resonance (and nonresonant S-wave), allowing  the improvement of the  $\phi_s$ accuracy and, in addition, further studies of the resonance  composition in the decay. 

These $\Bs\to\jpsi K^+K^-$ decays are separated into two $K^+K^-$ mass intervals. Those with $\m <1.05$~GeV are called low-mass and correspond to the region of the $\phi(1020)$ resonance, while those with $\m >1.05$~GeV are called high-mass. The high-mass region has not been analyzed for \CP violation before, allowing the measurement of \CP violation in several decay modes, including a vector-tensor final state,  $\jpsi f_2^\prime(1525)$. In the SM the phase $\phi_s$ is expected to be the same in all such modes. One important difference  from the previous low-mass analysis \cite{LHCb-PAPER-2014-059} is that modelling of the $\m$ distribution is included to distinguish different resonance and nonresonance contributions. In the previous low mass \CP-violation analysis only the $\phi(1020)$ resonance and an S-wave amplitude were considered.   This analysis follows very closely the analyses of \CP violation in $\Bs\to\jpsi\pi^+\pi^-$ decays \cite{LHCb-PAPER-2014-019} and in $\B^0\to\jpsi\pi^+\pi^-$ decays~\cite{LHCb-PAPER-2014-058}, and only significant changes with respect to those measurements are described in this paper. The analysis strategy is to fit the \CP-even and \CP-odd components in the decay width probability density functions that describe the interfering amplitudes in the particle and antiparticle decays. These fits are done as functions of the  $\Bs$ proper decay time and in a four-dimensional phase space including the three helicity angles characterizing the decay and $\m$.  Flavour tagging, described below, allows us to distinguish between initial $\Bs$ and $\Bsb$ states.

This paper is organized as follows.  Section~\ref{sec:Brate} describes the $\Bs$ proper-time dependent decay widths. Section~\ref{sec:Detector} gives a description of the detector and the associated simulations. Section~\ref{sec:Selection} contains the event selection procedure and the extracted signal yields. Section~\ref{sec:Resolution} shows the measurement of the proper-time resolution and efficiencies for the final state in the four-dimensional phase space. Section~\ref{sec:tagging} summarizes the identification of the initial flavour of the state, a process called flavour tagging. Section~\ref{sec:resonance} gives the masses and widths of resonant states that decay into $K^+K^-$, and the description of a model-independent S-wave parameterization. Section~\ref{sec:likelihood} describes the unbinned likelihood fit procedure used to determine the physics parameters, and presents the results of the fit, while Section~\ref{sec:systematic} discusses the systematic uncertainties. Finally, the results are summarized and combined with other measurements in Section~\ref{sec:conclusion}.

\section{\boldmath Decay rates for $\Bs$ and $\Bsb \to \jpsi K^+ K^-$}
\label{sec:Brate}
The total decay amplitude for a \Bs (\Bsb) meson at decay time equal to zero is taken to be the sum over individual $K^+K^-$ resonant transversity amplitudes \cite{Dighe:1995pd}, and one nonresonant amplitude, with each component labelled as ${A}_i$ ($\overline{A}_i$).  Because of the spin-1 $\jpsi$ in the final state, the three possible polarizations of the $\jpsi$ generate longitudinal ($0$), parallel ($\parallel$) and perpendicular ($\perp$) transversity amplitudes. When the $K^+K^-$ forms a spin-0 state the final system only has a longitudinal component. Each of these amplitudes is a  pure \CP eigenstate. By introducing the parameter $\lambda_i \equiv \frac{q}{p}\frac{\Ab_i}{A_i}$, relating \CP violation in the interference between mixing and decay associated with the state $i$, the total amplitudes ${\cal A}$ and $\cAb$ can be expressed as the sums of the individual $\BorBb$ amplitudes,  ${\cal A}=\sum A_i$ and $\cAb =\sum \frac{q}{p} \Ab_i =\sum \lambda_i A_i= \sum \eta_i |\lambda_i| e^{-i\phi_s^i} A_i$. The quantities $q$ and $p$ relate the mass to the flavour eigenstates \cite{Bigi:2000yz}. For each transversity state $i$ the \CP-violating phase $\phi_s^i\equiv -\arg(\eta_i\lambda_i)$~\cite{Aaij:2013oba}, with $\eta_i$ being the \CP eigenvalue of the state. Assuming that any possible \CP violation in the decay is the same for all amplitudes, then $\lambda\equiv \eta_i\lambda_i$ and $\phi_s\equiv -\arg(\lambda)$ are common.  The decay rates into the $\jpsi K^+K^-$ final state are\footnote{$|p/q|=1$ is used. The latest LHCb measurement determined $|p/q|^2=1.0039\pm0.0033$ \cite{LHCb-PAPER-2016-013}.}
\begin{eqnarray}\label{Eq-t}
\Gamma(t) \propto
 e^{-\Gs t}\left\{\frac{|\A|^2+|\cAb|^2}{2}\ch  + \frac{|\A|^2-|\cAb|^2}{2}\cs\right.\quad\quad\nonumber\\
- \left.\Real(\A^*\cAb)\sh  -  \Imag(\A^*\cAb)\sn\right\},
\end{eqnarray}
\begin{eqnarray}
\overline{\Gamma}(t) \propto
 e^{-\Gs t}\left\{\frac{|\A|^2+|\cAb|^2}{2}\ch  - \frac{|\A|^2-|\cAb|^2}{2}\cs\right.\quad\quad\nonumber\\
- \left.\Real(\A^*\cAb)\sh  +  \Imag(\A^*\cAb)\sn\right\},\label{Eqbar-t}
\end{eqnarray}
where $\DGs \equiv \GL-\GH$ is the decay width difference between the light and the heavy mass eigenstates, $\dms \equiv m_{\rm H}-m_{\rm L}$ is the mass difference, and $\Gs \equiv (\GL+\GH)/2$ is the average width.  The sensitivity to the phase $\phi_s$ is driven by the terms containing $\A^*\cAb$.

\begin{figure}[t]
  \centering
  \includegraphics{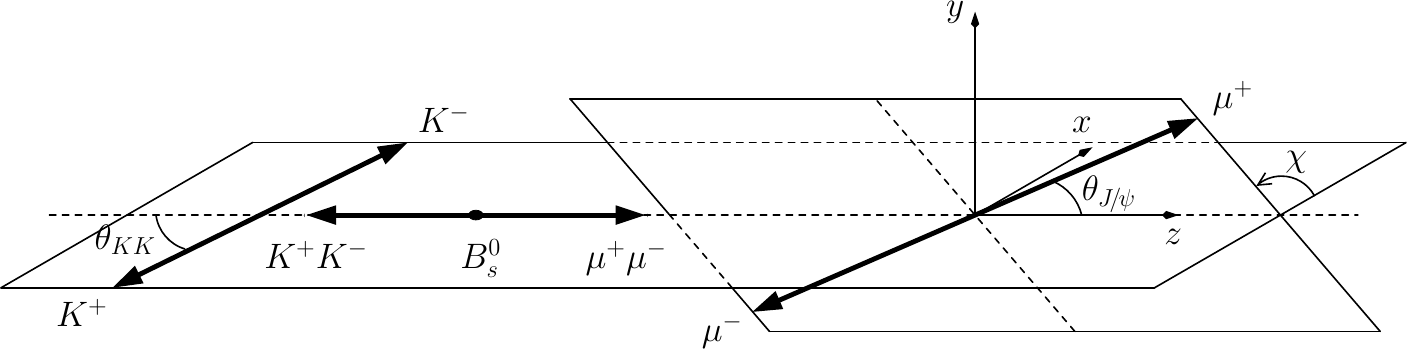}
  \caption{\small Definition of the helicity angles.}
  \label{fig:helicityAngles}
\end{figure}

For $\jpsi$ decays to $\mu^+\mu^-$ final states, these amplitudes are themselves functions of four variables: the $\Kp\Km$ invariant mass $\m$, and  three angular variables $\Omega\equiv (\cos\angpi, \cos\angmu, \chi)$, defined in the helicity basis. These consist of the angle $\angpi$ between the $K^+$ direction in the $K^+K^-$ rest frame with respect to the $K^+K^-$ direction in the $\Bs$ rest frame, the angle $\angmu$ between the $\mu^+$ direction in the $\jpsi$ rest frame with respect to the $\jpsi$ direction in the $\Bs$ rest frame,  
and the angle $\chi$ between the $\jpsi$ and $K^+K^-$ decay planes in the $\Bs$ rest frame \cite{Zhang:2012zk,Aaij:2013oba}.  These angles are shown pictorially in Fig.~\ref{fig:helicityAngles}. These definitions are the same for $\Bq$ and $\Bqb$, namely, using $\mu^+$ and $K^+$ to define the angles for both $\Bq$ and $\Bqb$ decays.
The explicit forms of $|{\cal A}(\m,\Omega)|^2$,  $|\cAb(\m,\Omega)|^2$, and $\A^*(\m,\Omega)\cAb(\m,\Omega)$ in Eqs.~(\ref{Eq-t}) and (\ref{Eqbar-t}) are given in Ref.~\cite{Zhang:2012zk}.

%

\section{Detector and simulation}
\label{sec:Detector}

The \lhcb detector~\cite{Alves:2008zz,LHCb-DP-2014-002} is a single-arm forward
spectrometer covering the \mbox{pseudorapidity} range $2<\eta <5$,
designed for the study of particles containing \bquark or \cquark
quarks. The detector includes a high-precision tracking system
consisting of a silicon-strip vertex detector surrounding the $pp$
interaction region, a large-area silicon-strip detector located
upstream of a dipole magnet with a bending power of about
$4{\mathrm{\,Tm}}$, and three stations of silicon-strip detectors and straw
drift tubes placed downstream of the magnet.
The tracking system provides a measurement of momentum, \ptot, of charged particles with
a relative uncertainty that varies from 0.5\% at low momentum to 1.0\% at 200\gev.
The minimum distance of a track to a primary vertex (PV), the impact parameter (IP), 
is measured with a resolution of $(15+29/\pt)\mum$,
where \pt is the component of the momentum transverse to the beam, in\,\gev.
Different types of charged hadrons are distinguished using information
from two ring-imaging Cherenkov (RICH) detectors. 
Photons, electrons and hadrons are identified by a calorimeter system consisting of
scintillating-pad and preshower detectors, an electromagnetic
calorimeter and a hadronic calorimeter. Muons are identified by a
system composed of alternating layers of iron and multiwire
proportional chambers.

The online event selection is performed by a trigger, 
which consists of a hardware stage, based on information from the calorimeter and muon
systems, followed by a software stage, which applies a full event
reconstruction. The software trigger is composed of two stages, the first of which performs a partial reconstruction and requires either a pair of well-reconstructed, oppositely charged muons having an invariant mass above 2.7\gev, or a single well-reconstructed muon with high \pt and large IP. The second stage  applies a full event reconstruction and for this analysis requires two opposite-sign muons to form a good-quality vertex that is well separated from all of the PVs, and to have an invariant mass within $\pm120$\mev of the known \jpsi mass~\cite{PDG2016}.

In the simulation, $pp$ collisions are generated using
\pythia8~\cite{Sjostrand:2006za,*Sjostrand:2007gs}. Decays of hadronic particles
are described by \evtgen~\cite{Lange:2001uf}, in which final-state
radiation is generated using \photos~\cite{Golonka:2005pn}. The
interaction of the generated particles with the detector, and its response,
are implemented using the \geant
toolkit~\cite{Allison:2006ve, *Agostinelli:2002hh} as described in
Ref.~\cite{LHCb-PROC-2011-006}. The simulation covers the full $K^+K^-$ mass range.

\section{Event selection and signal yield extraction}
\label{sec:Selection}
A $\Bs$ candidate is reconstructed by combining a $\jpsi \to \mu^+\mu^-$ candidate with two kaons of opposite charge. 
The offline selection uses a loose preselection, followed by a multivariate classifier based on a Gradient Boosted Decision Tree (BDTG)~\cite{Breiman}.

In the preselection, the \jpsi candidates are formed from two oppositely charged particles with $\pt$ greater than 550$\mev$, identified as muons and consistent with originating from a common vertex but inconsistent with originating from any PV. The invariant mass of the $\mumu$ pair is required to be within $[-48, +43]$~MeV of the known \jpsi mass~\cite{PDG2016}, corresponding to a window of about $\pm3$ times the mass resolution. The asymmetry in the cut values is due to the radiative tail. The two muons are subsequently kinematically constrained to the known $\jpsi$ mass. Kaon candidates are required to be positively identified in the RICH detectors, to have $\pt$ greater than 250\mev, and the scalar sum of the two transverse momenta, $\pt(K^+)+\pt(K^-)$, must be larger than 900\mev.

The four tracks from a $\Bs$ candidate decay must originate from a common vertex with a good fit $\chi^2$ and have a decay time greater than 0.3\ps. 
Each $\BorBb$ candidate is assigned to a PV for which it has the smallest $\chisqip$,  defined as the difference in the $\chisq$ of the vertex fit for a given PV reconstructed with and without the considered particle.
The angle between the momentum vector of the $\Bs$ decay candidates
and the vector formed from the positions of the PV and
the decay vertex (pointing angle) is required to be less than $2.5^{\circ}$. 

Events are filtered with a BDTG to further suppress the combinatorial background. The BDTG uses six variables: $\pt(K^+)+\pt(K^-)$; the vertex-fit $\chi^2$, pointing angle, $\chisqip$, and $\pt$ of the $\BorBb$ candidates; and the smaller of the DLL($\mu-\pi$) for the two muons, where DLL($\mu-\pi$) is the difference in the logarithms of the likelihood values from the particle identification systems~\cite{LHCb-DP-2012-003} for the muon and pion hypotheses. The BDTG is trained on a simulated sample of 0.7 million reconstructed $\Bsjpsikk$  signal events, with the final-state particles generated uniformly in phase space assuming unpolarized $J/\psi \rightarrow \mu^+\mu^-$ decays, and a background data sample from the sideband $5516<m(J/\psi K^+K^-)< 5616$\,\mev. Separate samples are used to train and test the BDTG. The 
BDTG and particle identification (PID) requirements for the kaons are chosen to maximize the signal significance multiplied by the square root of the purity, $S/\sqrt{S+B}\times\sqrt{S/(S+B)}$, for candidates with $\m>1.05$\gev, where $S$ and $B$ are the numbers of  signal and  background candidate combinations, respectively. This figure of merit optimizes the total uncertainty including both statistical and background systematic errors.
\begin{figure}[btp]
\centering
\includegraphics[width=0.5\textwidth]{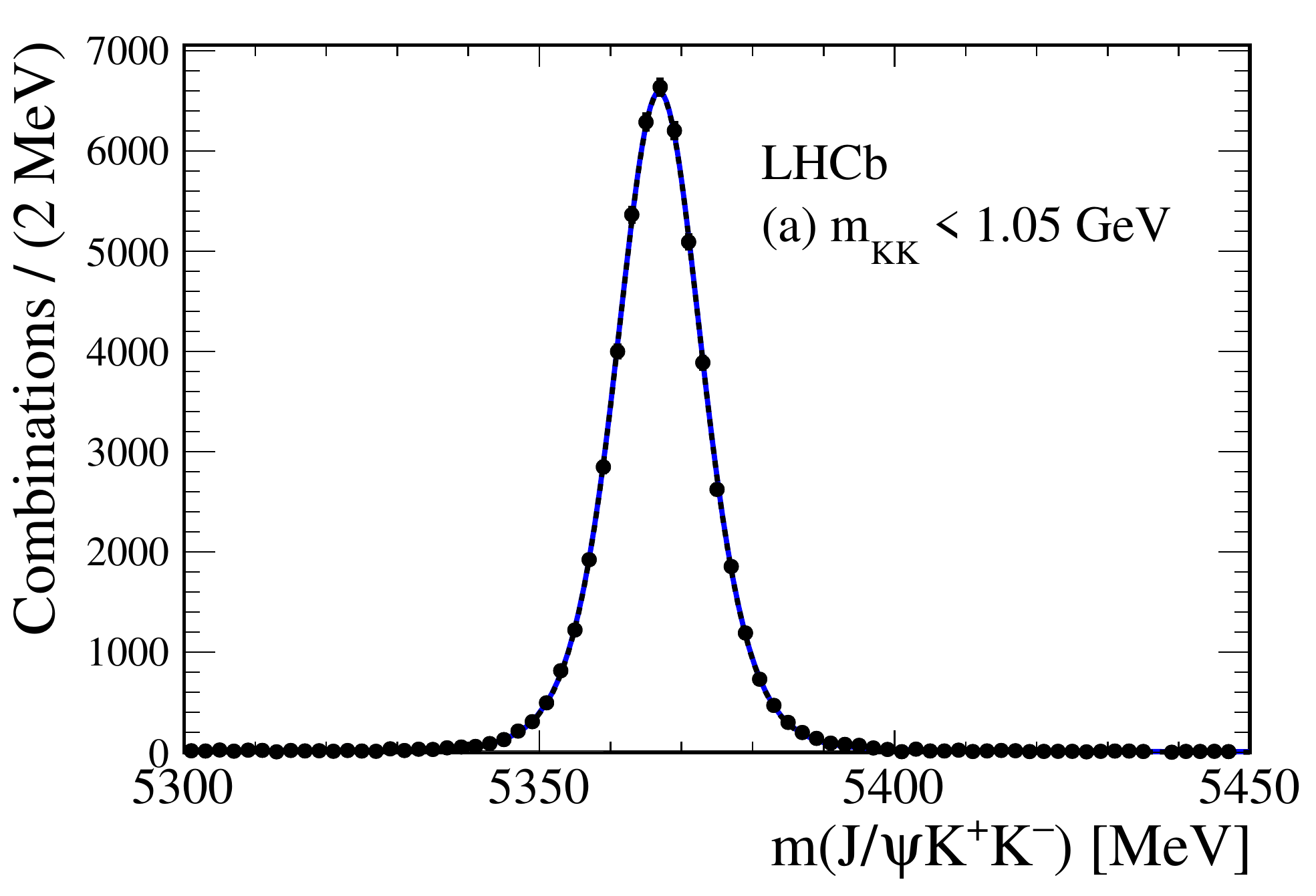}%
\includegraphics[width=0.5\textwidth]{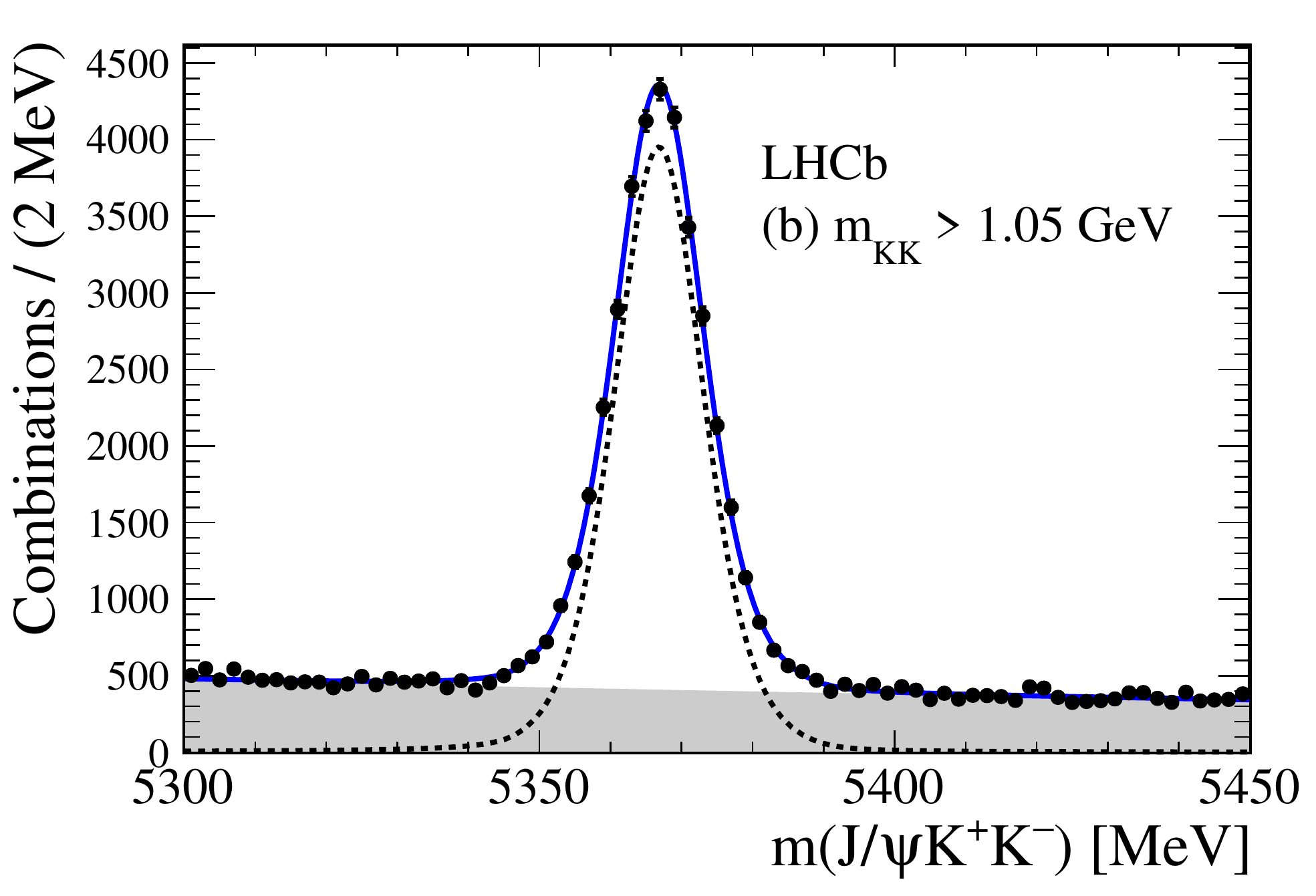}
\caption{\small Fits to invariant mass distributions of $\jpsi K^+K^-$ combinations after subtraction of the two reflection backgrounds for (a) $\m<1.05$\gev and (b) $\m>1.05$\gev. Total fits are shown by solid (blue) lines, the signal by dashed (black) lines, and the combinatorial background by darkened regions. Note that the combinatorial background in (a) is too small to be easily visible.}
\label{massfit-sub}
\end{figure}

In addition to the expected
combinatorial background, studies of the data in sidebands of the $m(\jpsi \Kp \Km)$ spectrum show contributions from approximately  $8700$ ($430$) $\BzbJpsiKpi$ and  $10\,700$
 ($800$) $\LbJpsipK$ decays at $\m$ greater (less) than 1.05\gev, where the $\pip$ in the former or $p$ in the latter is misidentified as a \Kp. In order to avoid dealing with correlations between the angular variables and $m(\jpsi K^+K^-)$, the contributions from these reflection backgrounds are statistically subtracted by adding to the data simulated events of these decays with negative weights. 
These weights are chosen so that the distributions of the relevant variables used in the overall fit (see below) describe the background distributions both in normalization and shapes. The simulation uses amplitude models derived from data for $\Bdb\to \jpsi \Km \pip$~\cite{Chilikin:2014bkk} and $\Lb \to \jpsi p \Km$ decays~\cite{LHCb-PAPER-2015-029}. 

\begin{figure}[btp]
\centering
\includegraphics[width=0.7\textwidth]{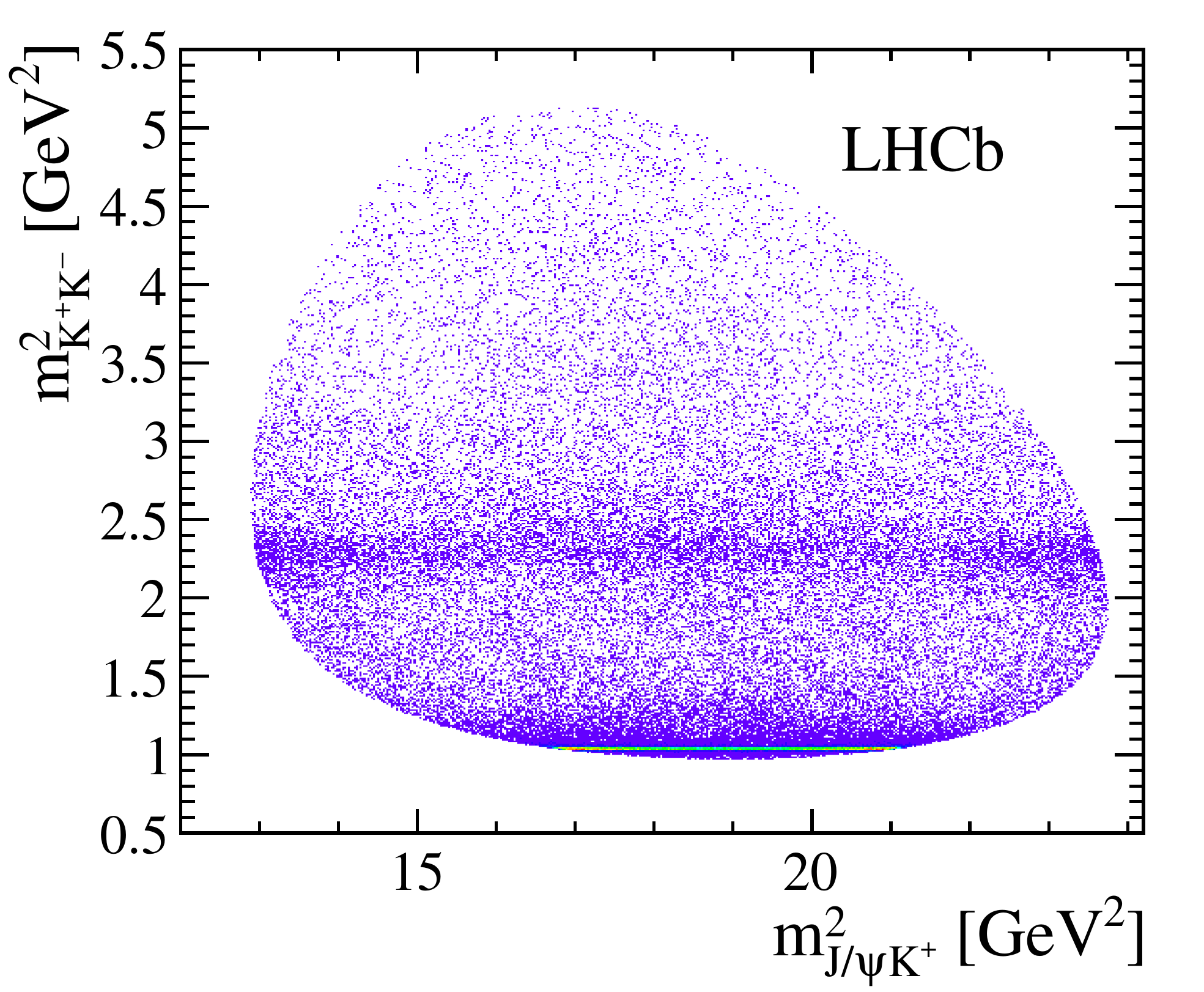}
\caption{\small Invariant mass squared of $K^+K^-$ versus $\jpsi K^+$ for $\Bs\to\jpsi K^+K^-$ candidates within $\pm15$\mev of the $\Bs$ mass peak. The high intensity $\phi(1020)$ resonance band is shown with a line (light green). }
\label{dlz}
\end{figure}

The invariant mass of the selected $\jpsi K^+K^-$ combinations, separated into samples for $\m$ below or above 1.05\gev, are shown in Fig.~\ref{massfit-sub}, where the expected reflection backgrounds are subtracted using simulation. The combinatorial background is modelled with an exponential function and the $\Bs$ signal shape is parameterized by a double-sided Hypatia function~\cite{Santos:2013gra}, where the signal radiative tail parameters are fixed to values obtained from simulation.  In total, $53\,440\pm240$ and $33\,200\pm240$ signal candidates are found for the low and high $\m$ intervals, respectively. 
Figure~\ref{dlz} shows the Dalitz plot distribution of  $m^2_{K^+K^-}$  versus  $m^2_{\jpsi K^+}$ for $\Bs\to\jpsi K^+K^-$ candidates within $\pm$15\mev of the $\Bs$ mass peak. Clear resonant contributions from $\phi(1020)$ and $f_2^\prime(1525)$ mesons are seen, but no exotic $\jpsi K^+$ resonance  is observed.

\section{Detector resolution and efficiency}
\label{sec:Resolution}

The resolution on the decay time is determined
with the same method as
described in Ref.~\cite{LHCb-PAPER-2014-059} by
using a large sample of  prompt $\jpsi\Kp\Km$ combinations produced directly in the $pp$ interactions. These
events are selected using  ${\jpsi \to \mu^+\mu^-}$ decays via a prescaled
trigger that does not impose any requirements on the separation of the \jpsi
from the PV. The \jpsi candidates are combined with two oppositely charged tracks that are identified
as kaons, using a similar selection as for the signal decay, without a  decay-time requirement. The
resolution function, $T(t-\hat{t}\,|\,\delta_t)$, where $\hat{t}$ is the true decay time, is a sum of three Gaussian functions with a common mean, and separate widths. To implement the resolution model each of the three widths are  given by $S_i\cdot(\delta_t+\sigma_t^0)$, where $S_i$ is scale factor for the $i$th Gaussian, $\delta_t$ is an estimated per-candidate decay-time error and $\sigma_t^0$ is a constant parameter. The parameters of the resolution model  are determined by using a
maximum likelihood fit to the unbinned decay time and $\delta_t$ distributions
of the prompt $\jpsi\Kp\Km$ combinations, using a $\delta$ function to represent the prompt component summed with two exponential functions for long-lived backgrounds; these are convolved with the resolution function.
Taking into account the $\delta_t$ distribution of the $\BorBb$ signal, the average effective resolution is found to be $44.7$\fs.

\begin{figure}[t]
\centering{
   \includegraphics*[width=0.45\textwidth]{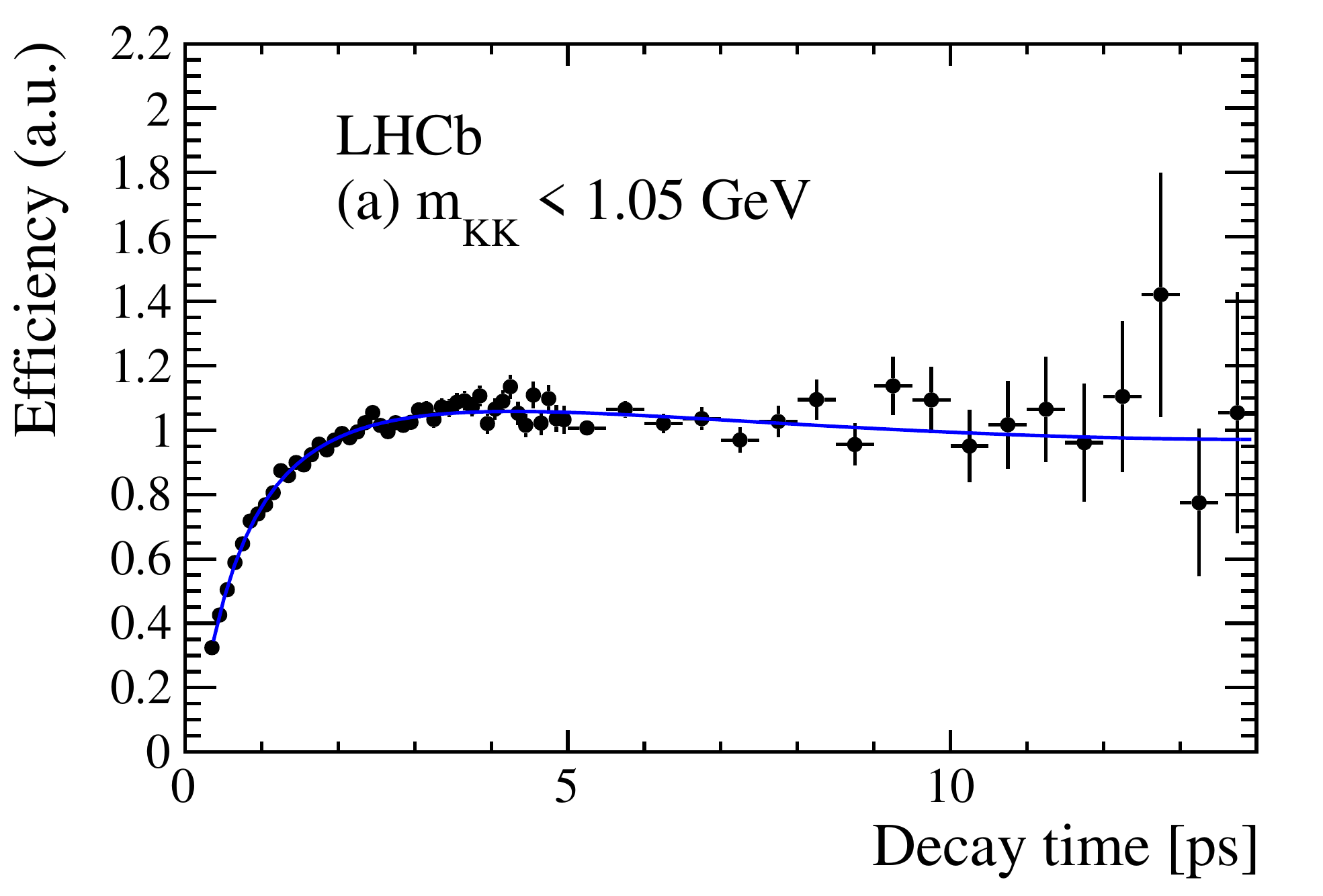}%
   \includegraphics*[width=0.45\textwidth]{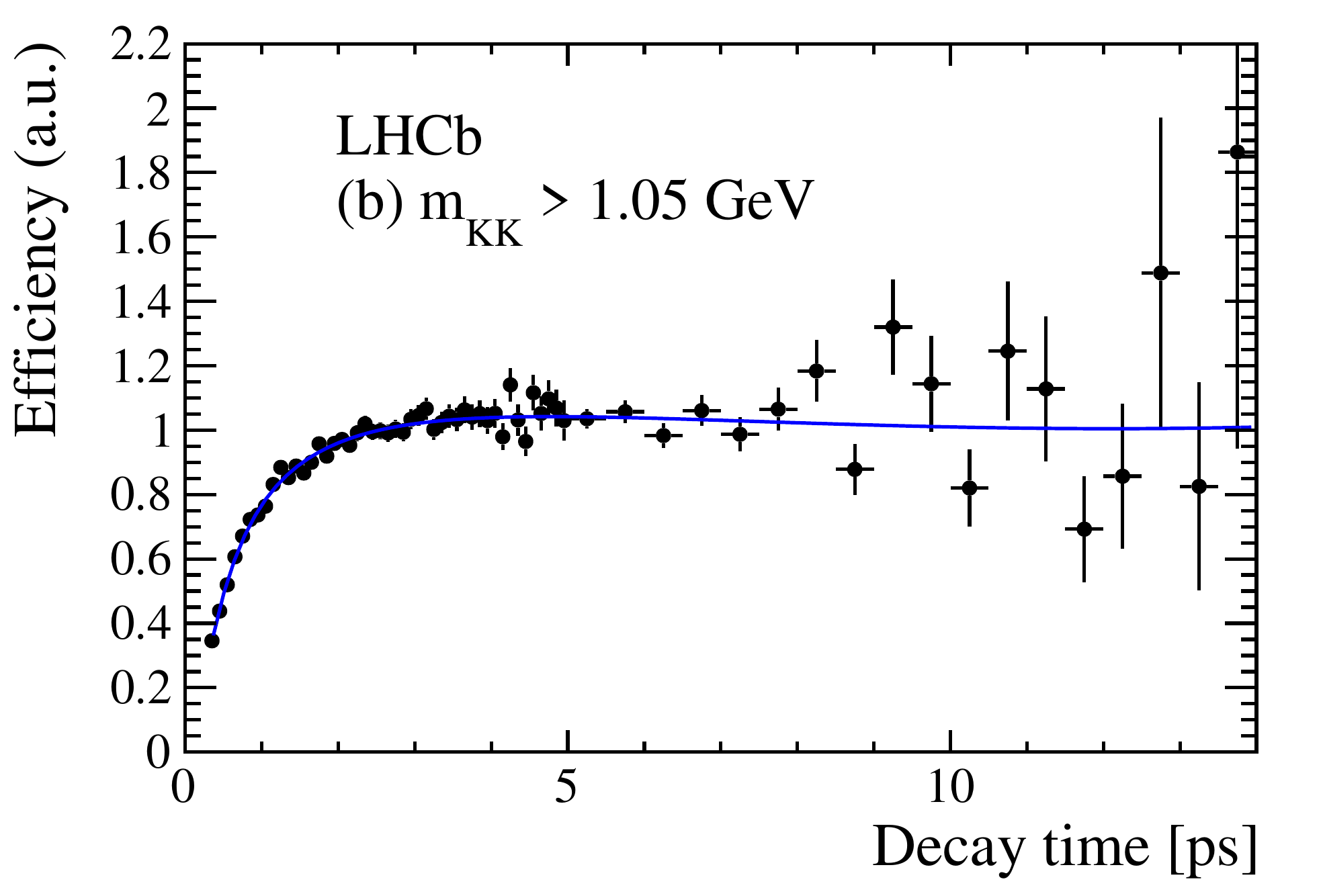}
    }
  \caption{\small Scaled decay-time efficiency $\varepsilon_{\rm data}^{\Bs}(t)$ in arbitrary units (a.u.) for (a) the $\phi(1020)$ region and (b) the high-mass region. }
  \label{fig:bs_data}
\end{figure}

The reconstruction efficiency is not constant as a function of decay time due to
displacement requirements made on the $\jpsi$ candidates in the trigger and offline selections. The
efficiency is determined using the control channel $\Bd \to \jpsi K^{*}(892)^0$,
with $K^{*}(892)^0 \to \Kp\pi^-$, which is known to have
a purely exponential decay-time distribution with $\tau_{\Bd} = 1.520 \pm 0.004\ps$~\cite{PDG2016}. The selection efficiency is calculated as
\begin{equation}\label{eqn:acc}
	\varepsilon_{\rm data}^{\Bs}(t) = \varepsilon_{\rm data}^{\Bd}(t) \times \frac{\varepsilon_{\rm sim}^{\Bs}(t)}{\varepsilon_{\rm sim}^{\Bd}(t)},
\end{equation}
where $\varepsilon_{\rm data}^{\Bd}(t)$ is the efficiency of the control channel and
$\varepsilon_{\rm sim}^{\Bs}(t)/\varepsilon_{\rm sim}^{\Bd}(t)$ is the ratio of efficiencies of the
simulated signal and control mode after the full trigger and selection chain has been applied.
This correction accounts for the small differences in the kinematics between
the signal and control mode. The details of the method are explained in Ref.~\cite{LHCb-PAPER-2014-019}. The decay-time efficiencies for the two $\m$ intervals are shown in Fig.~\ref{fig:bs_data}.

The efficiency as a function of the $\Bs\to\jpsi K^+K^-$ helicity angles and the $K^+K^-$ invariant mass is not
uniform due to the forward geometry of the LHCb detector and the
requirements imposed on the final-state
particle momenta. The four-dimensional efficiency, $\varepsilon(\m,\Omega)$,
is determined using simulated events that are subjected to
the same trigger and selection criteria as the data.

\begin{figure}[t]
  \begin{center}
     \includegraphics[width=0.5\textwidth]{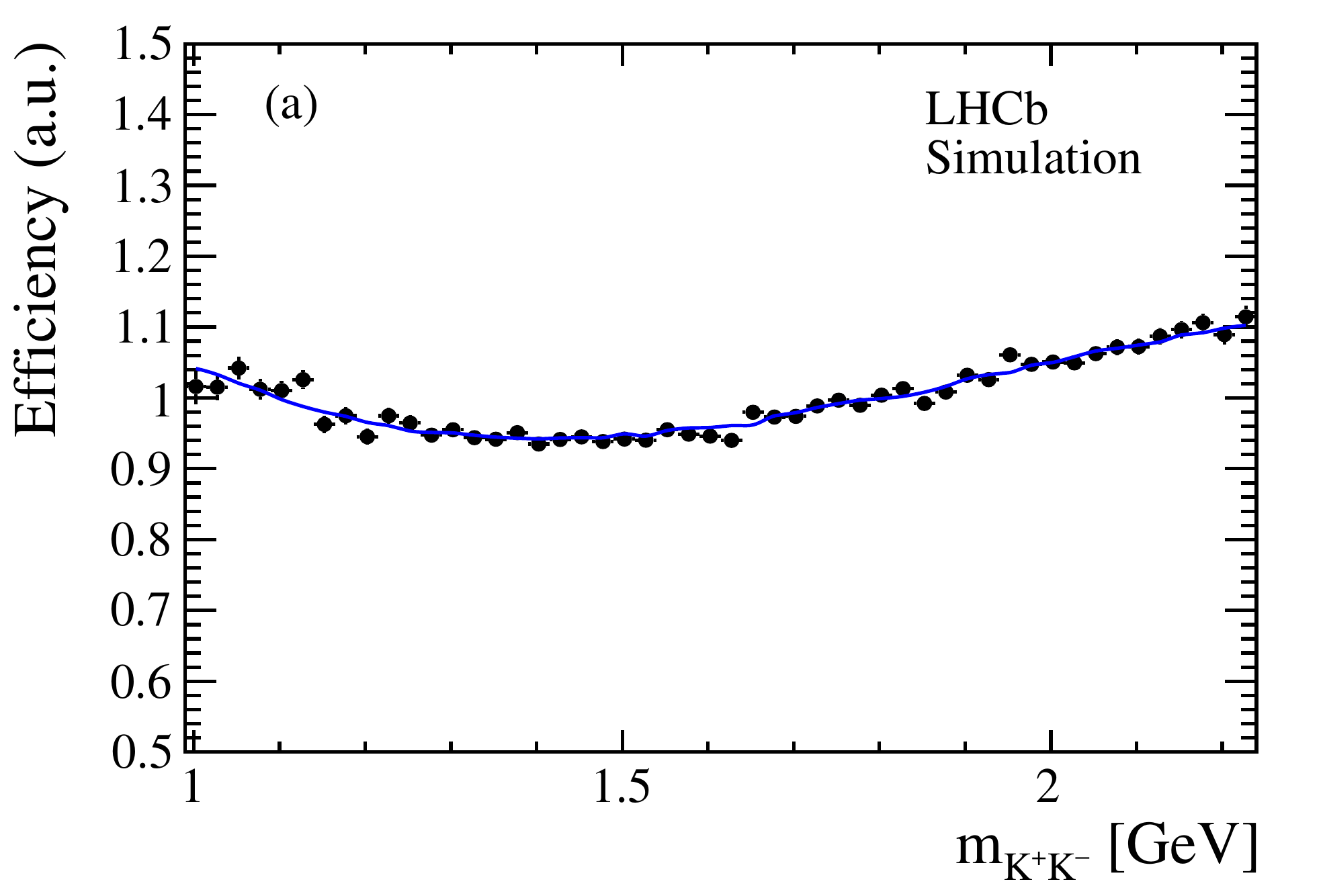}%
     \includegraphics[width=0.5\textwidth]{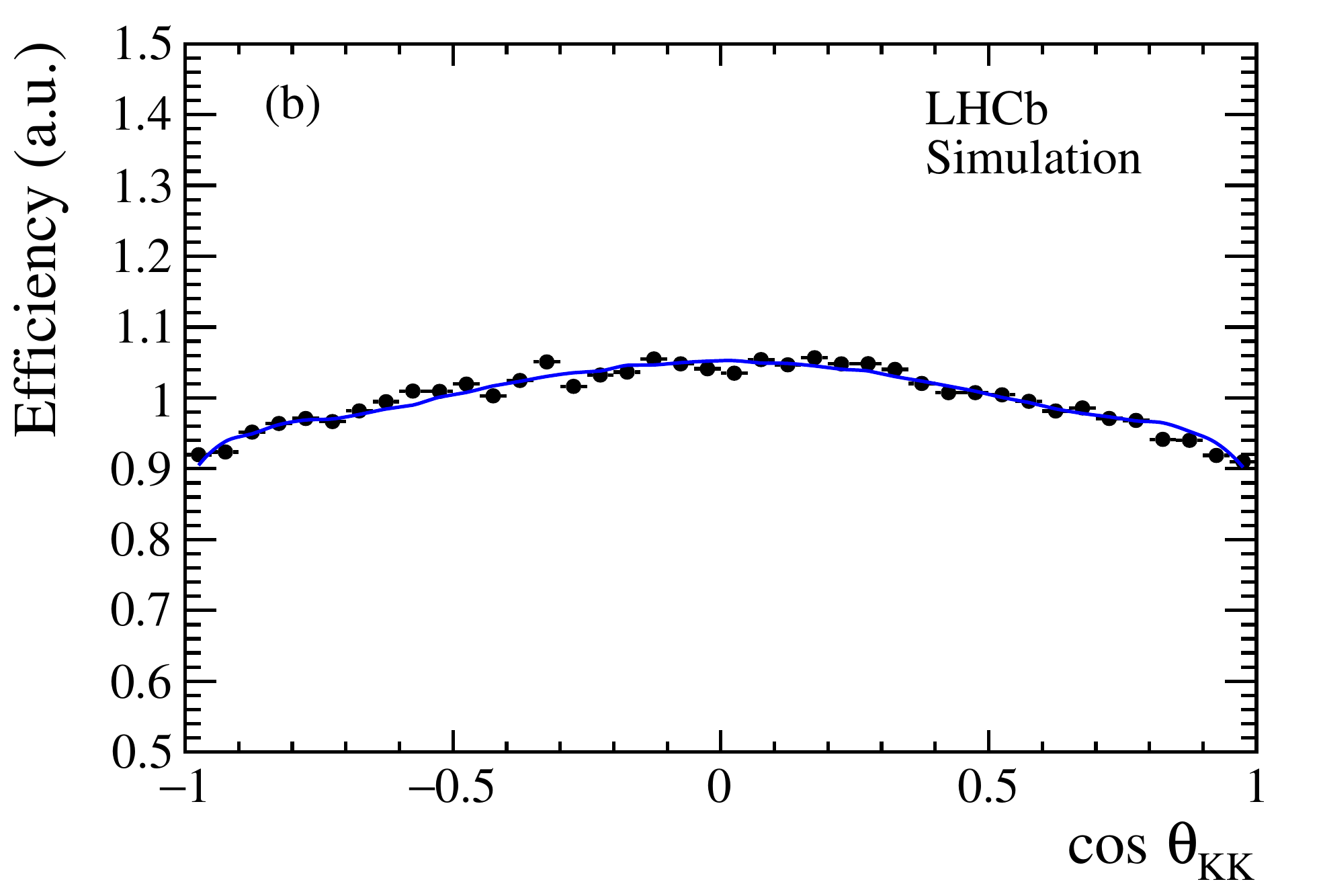}\\
      \includegraphics[width=0.5\textwidth]{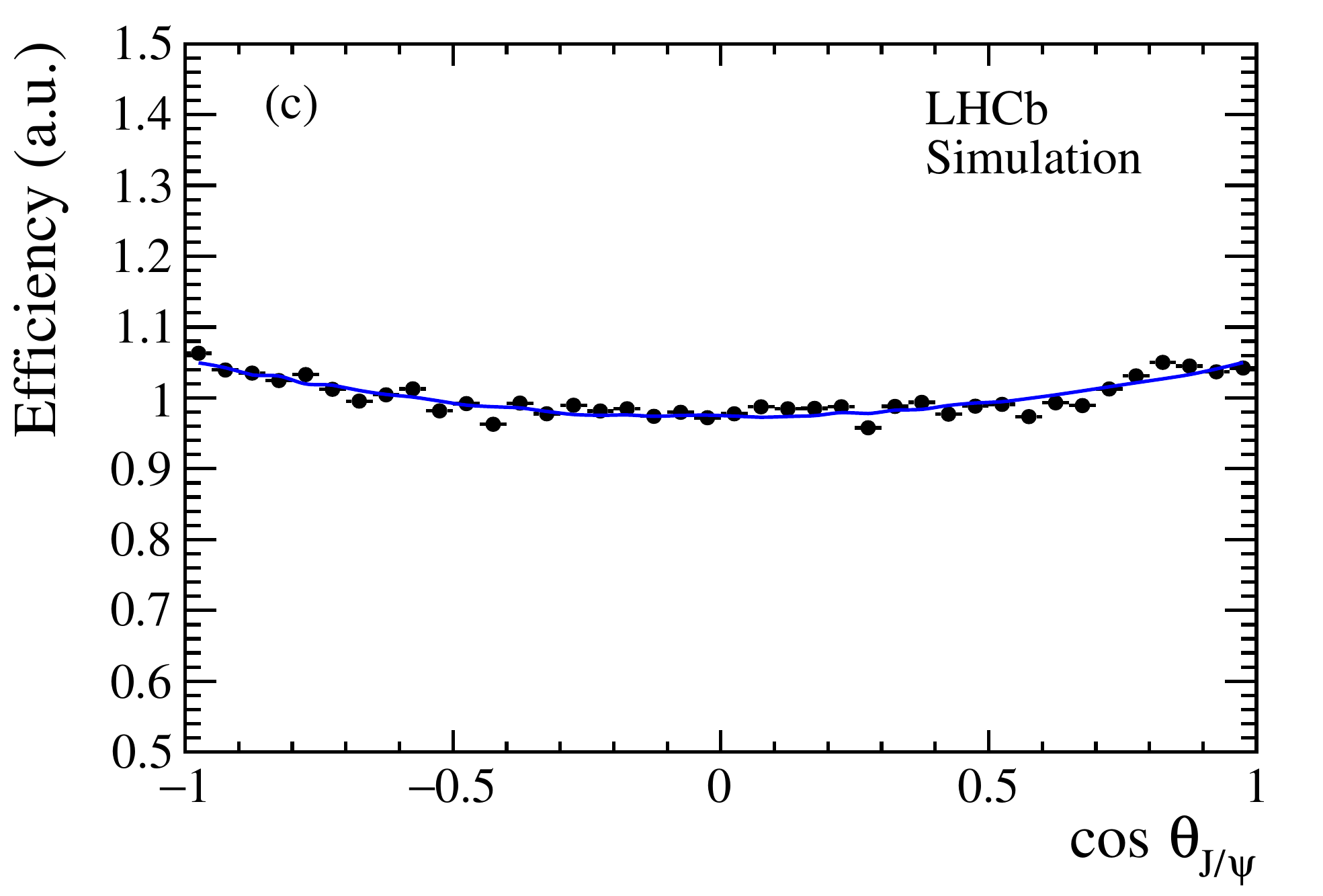}%
     \includegraphics[width=0.5\textwidth]{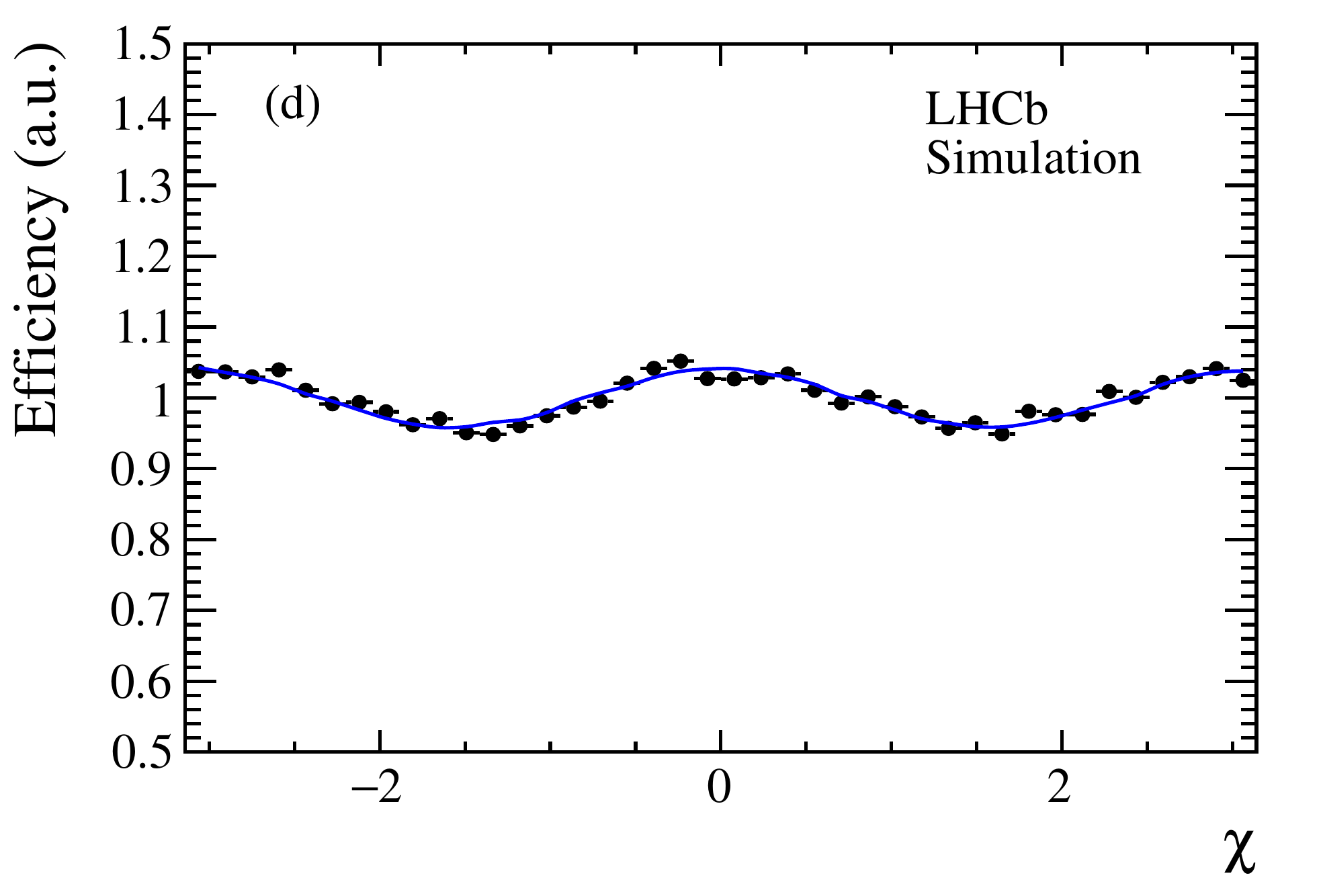}\\
    \caption{\small Efficiencies projected onto  (a) $\m$, (b) $\cos \angpi$, (c) $\cos \angmu$ and (d) $\chi$ in arbitrary units (a.u.), obtained from  simulation of  $\Bsjpsikk$ phase-space decays (points with error bars), while the curves show the parameterization from  the efficiency model.}  \label{fig:acc}
  \end{center}
\end{figure}

The efficiency is parameterized by 
\begin{equation}\label{EqEff}
\epsilon(\m,\Omega) = \sum_{a,b,c,d}\epsilon^{abcd}{\cal{P}}_a(\cos\angpi)Y_{bc}(\angmu,\chi){\cal{P}}_d\left(2\frac{\m-\m^{\rm min}}{\m^{\rm max}-\m^{\rm min}}-1\right),
\end{equation}
where ${\cal{P}}_a$ and ${\cal{P}}_d$ are Legendre polynomials, $Y_{bc}$ are spherical harmonics, and $\m^{\rm min}=2 m_{K^+}$ and $\m^{\rm max}=m_{\Bs}-m_{\jpsi}$ are the minimum and maximum allowed values for $\m$, respectively. The $Y_{bc}$ are complex functions. To ensure that the efficiency function is real, we set $\epsilon^{abcd}=-\epsilon^{ab(-c)d}$. The values of $\epsilon^{abcd}$ are determined by summing over  the fully simulated phase-space events 
\begin{equation}\label{eq:effco}
\epsilon^{abcd}=\frac{1}{\sum_i w_i}\sum_i w_i \frac{2a+1}{2}\frac{2d+1}{2}{\cal{P}}_a(\cos {\angpi}\!_{,i})Y^*_{bc}({\angmu}\!_{,i},\chi_i){\cal{P}}_d\left(2\frac{{\m}_{,i}-\m^{\rm min}}{\m^{\rm max}-\m^{\rm min}}-1\right)\frac{1}{g_i},
\end{equation}
where the weights $w_i$ account for corrections of PID and tracking efficiencies, and $g_i=P^i_{R}P^i_{B}$ is the value of the phase-space probability density for event $i$ with $P_{R}$ being the momentum of either of the two hadrons in the dihadron rest frame and $P_{B}$ the momentum of the $\jpsi$ in the $\BorBb$ rest frame. This approach allows the description of multidimensional correlations without assuming factorization. In practice, the sum is over a finite number of terms ($a\leq10$, $b\leq8$, $-2\leq c\leq2$, $d\leq8$) and only coefficients with a statistical significance larger than three standard deviations $(\sigma)$ from zero are retained. The number of events in the  simulated signal sample is about 20 times of that observed in data.  Since a symmetric $K^+$ and $K^-$ efficiency is used, $a$ and $b+c$ must be even numbers. Projections of the efficiency integrated over other variables are shown in Fig.~\ref{fig:acc}. The modelling functions describe well the simulated data.  Since \chisqip is not used as a variable in the selection for the two hadrons, the efficiency is quite uniform over all the four variables varying only by about $\pm10$\%. (A dedicated simulation of $\jpsi\phi(1020)$ decays is used to determine the efficiency in the region of $\m<1.05$\gev, in order to have a large enough sample for an accurate determination.)

\section{Flavour tagging}
\label{sec:tagging}

The \Bs candidate flavour at production is determined using two independent classes of
flavour-tagging algorithms, the opposite-side (OS) tagger~\cite{LHCb-PAPER-2011-027} and the same-side kaon (SSK)
tagger~\cite{LHCb-PAPER-2015-056}, which exploit specific features of the production of $\bquark \bquarkbar$
quark pairs in $\proton\proton$ collisions, and their subsequent hadronisation.
Each tagging algorithm provides a tag decision and a mistag probability.
The tag decision, $\mathfrak{q}$, is $+1$, $-1$, or $0$, if the signal meson is tagged as $\Bs$, $\Bsb$, or is untagged,
respectively. The fraction of candidates in the sample with a nonzero tagging decision gives
the efficiency of the tagger, $\varepsilon_{\rm tag}$.
The mistag probability, $\eta$, is estimated event by event, and
represents the probability that the algorithm assigns a wrong tag decision to the candidate; it
is calibrated using data samples of several flavour-specific $\Bd$, $\Bu$, $\Bs$ and $B_{s2}^{*0}$ \cite{LHCb-PAPER-2015-056} decays to obtain the corrected mistag probability, ${\omega}$, for an initial
 \BorBb meson, and separately obtain $\overline{\omega}$ for an initial \Bsb meson. A linear relationship between $\eta$
and $\optbar{\omega}$ is used for the calibration. When candidates are tagged by both the OS and the SSK algorithms, a combined tag decision and a wrong-tag probability are given by the algorithm defined in Ref.~\cite{LHCb-PAPER-2011-027} and extended to include SSK tags. This combined algorithm is implemented in the overall fit.
The effective tagging power is
given by $\varepsilon_{\rm tag}\mean{(1-2\omega)^2}$
and for the
combined taggers in the $\Bsjpsikk$ signal sample is $(3.82 \pm 0.13 \pm 0.12)\%$.
Whenever two uncertainties are quoted in this paper, the first is statistical and the second is systematic.

\section{Resonance contributions}
\label{sec:resonance}
The entire $K^+K^-$ mass spectrum is fitted by including the resonance contributions previously found in the time-integrated amplitude analysis using 1\invfb of integrated luminosity~\cite{LHCb-PAPER-2012-040}, except for the unconfirmed $f_2(1640)$ state. They are shown in Table~\ref{tab:res} and are described by Breit-Wigner amplitudes.  
The S-wave amplitude ${\cal S}(\m)=c(\m)+i s(\m)$ is described in a model-independent way, making 
no assumptions about  its $f_0$ meson composition, or about the form of any S-wave nonresonant terms. 
Explicitly, two real parameters $c^k=c(\m^k)$ and $s^k=s(\m^k)$ are introduced to define the total S-wave amplitude at each of a set of invariant mass values $\m=\m^k$ $(k=1,..,N_s)$. Third-order spline interpolations are used to define $c(\m)$ and $s(\m)$ between these points of $\m^k$. The $c^k$ and $s^k$ values are treated as model-independent parameters, and are determined by a fit to the data. In total $N_s=13$ knots are chosen at $\m =(1.01, 1.03, 1.05, 1.10, 1.40, 1.50, 1.65, 1.70,
1.75, 1.80, 1.90, 2.1, 2.269)$ \gev. The S-wave amplitude is proportional to momentum $P_B$~\cite{Zhang:2012zk}; at the last point since $P_B=0$, the amplitude is zero \cite{Zhang:2012zk}.

\begin{table}[bt]
\centering
\caption{\small Breit-Wigner resonance parameters.}\label{tab:res}
\vspace{-0.2cm}
\begin{tabular}{cccc}
 Resonance &Mass (MeV) & Width (MeV) & Source \\
 \hline
$\phi(1020)$ & $1019.461\pm0.019$ &$4.266\pm0.031$&PDG~\cite{PDG2016}\\
$f_2(1270)$ & $1275.5\pm 0.8$ & $186.7^{+2.2}_{-2.5}$&PDG~\cite{PDG2016} \\
$f_2'(1525)$ &\multicolumn{3}{c}{Varied in fits}\\
$\phi(1680)$& $1689\pm 12$ &$211\pm 24$&\belle~\cite{Shen:2009zze}\\
$f_2(1750)$ & $1737\pm 9$\,\,\, & $151\pm 33$ &\belle~\cite{Abe:2003vn}\\
$f_2(1950)$ & $1980\pm 14$ & $297\pm 13$&\belle~\cite{Abe:2003vn} \\
\end{tabular}\label{tab:resparam}
\end{table}

To describe the $\m$ dependence for each resonance $R$, the formula of Eq.~(18) in Ref.~\cite{Zhang:2012zk} is modified by changing $\left(\frac{P_R}{m_{KK}}\right)^{L_R}$ to $\left(\frac{P_R}{m_{0}}\right)^{L_R}$, where  $P_R$ is the momentum of either of the two hadrons in the dihadron rest frame, $m_{0}$ is the mass of
resonance $R$, and $L_R$ the orbital angular momentum in the $\Kp\Km$ decay, and thus corresponds to the  resonance's spin.  This change modifies the lineshape of resonances with spin greater than zero. The original formula followed the convention from the Belle collaboration~\cite{Mizuk:2008me} and was used in two LHCb publications~\cite{LHCb-PAPER-2012-040,LHCb-PAPER-2014-058}, while the new one follows the convention of PDG/\evtgen, and was used in analyzing $\Lb\to\jpsi p K^-$ decays~\cite{LHCb-PAPER-2015-029}. 
\section{Maximum likelihood fit}
\label{sec:likelihood}

The physics parameters are determined from a weighted maximum likelihood fit of a signal-only probability density function (PDF) to the five-dimensional distributions of  \Bs and \Bsb decay time, $\m$ and helicity angles. The negative log-likelihood function to be minimized is given by 

\begin{equation}
-\ln {\cal L} = - \alpha  \sum_i W_i \ln ({\cal PDF}),
\end{equation}  
where $i$ runs over all event candidates,  $W_i$ is the \sWeight computed using $m(J/\psi K^+K^-)$ as the discriminating variable \cite{Pivk:2004ty,*Xie:2009rka} and the factor  
$\alpha \equiv \sum_i W_i / \sum_i W_i^2$ is a constant factor accounting for the effect of the background subtraction on the statistical uncertainty.  The \sWeight{s} are determined by separate fits in four $|\cos\angmu|$ bins for the event candidates. 
The PDF is given by ${\cal PDF}={\cal F}/\int {\cal F} {\rm d}t\, {\rm d}\m\, {\rm d}\Omega$, where ${\cal F}$ is
\begin{equation}
{\cal F}(t,\m,\Omega,\mathfrak{q}\,|\,\eta,\delta_t)= \left[{\cal R}(\hat{t},\m,\Omega,\mathfrak{q}\,|\,\eta) \otimes T(t-\hat{t}\,|\,\delta_t)\right] \cdot \varepsilon_{\rm data}^{\Bs}(t)  \cdot \varepsilon(\m,\Omega), 
\end{equation}
with 
\begin{align}\label{eq:R}
{\cal R}(\hat{t},\m,\Omega,\mathfrak{q}\,|\,\eta) =&\frac{1}{1+|\mathfrak{q}|}\left[\left[1+\mathfrak{q}\left(1-2\omega(\eta)\right)\right]\Gamma(\hat{t},\m,\Omega)\right.\nonumber\\
&\left. +\left[1-\mathfrak{q}\left(1-2\bar{\omega}(\eta)\right)\right]\frac{1+A_{\rm P}}{1-A_{\rm P}}\bar{\Gamma}(\hat{t},\m,\Omega)\right],
\end{align}
where $\hat{t}$ is the true decay time, {\kern 0.18em\optbar{\kern -0.05em \Gamma}{}\xspace}  is defined in Eqs.~(\ref{Eq-t}) and~(\ref{Eqbar-t}), and  $A_{\rm P}=(1.09\pm2.69)\%$ is the LHCb measured production asymmetry of \Bs and \Bsb mesons~\cite{LHCb-PAPER-2014-042,*Aaij:2017mso}.

To obtain a measurement that is independent of the previous publication that used mainly $\jpsi\phi(1020)$ decays~\cite{LHCb-PAPER-2014-059}, two different sets of fit parameters ($\phi_s$, $|\lambda|$, \Gs, \DGs)$^{\rm L,H}$ are used to account for the low ($\rm L$) and high ($\rm H$) $\m$ regions. Simulated pseudoexperiments show that this configuration removes the correlation for these parameters between the two regions. A simultaneous fit to the two samples is performed by constructing the log-likelihood as the sum of that computed from the $\rm L$ and $\rm H$ events. The shared parameters are all the resonance amplitudes and phases, and  $\dms$, which is freely varied in the fit.  In the nominal fit configuration, \CP violation is assumed to be the same for all the transversity states. In total 69 free parameters are used in the nominal fit.

The \Bs decay observables resulting from the fit for the high $\m$ region are listed in Table~\ref{tab:fit}.  The measurements for these parameters and $\dms$ in the $\phi(1020)$ region are consistent with the reported values in Ref.~\cite{LHCb-PAPER-2014-059} within 1.4$\sigma$, taking into account the overlap between the two samples used. In addition, good agreement is also found for the S-wave phase. The fit gives $\dms=17.783\pm0.049\stat$\invps from the full $\m$ region, which is consistent with the most precise measurement $17.768\pm0.023\pm0.006$\invps from LHCb in $\Bs\to\Dsm\pi^+$ decays~\cite{LHCb-PAPER-2013-006}. The value of $|\lambda|$ is consistent with unity, thus giving no indication of any direct \CP violation in the decay amplitude.

\begin{table}[b]
\centering
\caption{\small Fit results for the \Bs decay observables in the high $\m$ region.}\label{tab:fit}
\begin{tabular}{lc}
Parameter & Value\\\hline
\Gs [\invps] &  $0.650\pm0.006\pm0.004$\\
\DGs [\invps] & $0.066\pm0.018\pm0.010$\\
$\phi_s$ [\,mrad\,]& $119\pm107\pm34$\\
$|\lambda|$&  $0.994\pm0.018\pm0.006$\\
\end{tabular}
\end{table}

While a complete description of the $\Bsb\to\jpsi K^+K^-$ decay is given in terms of the fitted amplitudes and phases, knowledge of the contribution of each component can be
summarized by the fit fraction, $FF_i$, defined as the integral of the squared amplitude of each resonance over the phase space divided by the integral of the entire signal function over the same area, as given in Eq.~\ref{eq:ff}

\begin{equation}
\label{eq:ff}
FF_i = \int{ |A_i|^2 d\m d\Omega}/\int {|{\cal A}|^2 d\m d\Omega}.
\end{equation}
The sum of the fit fractions is not necessarily unity due to the potential presence of interference between two resonances.

\begin{figure}[t]
\centering
\includegraphics[width=0.4\textwidth]{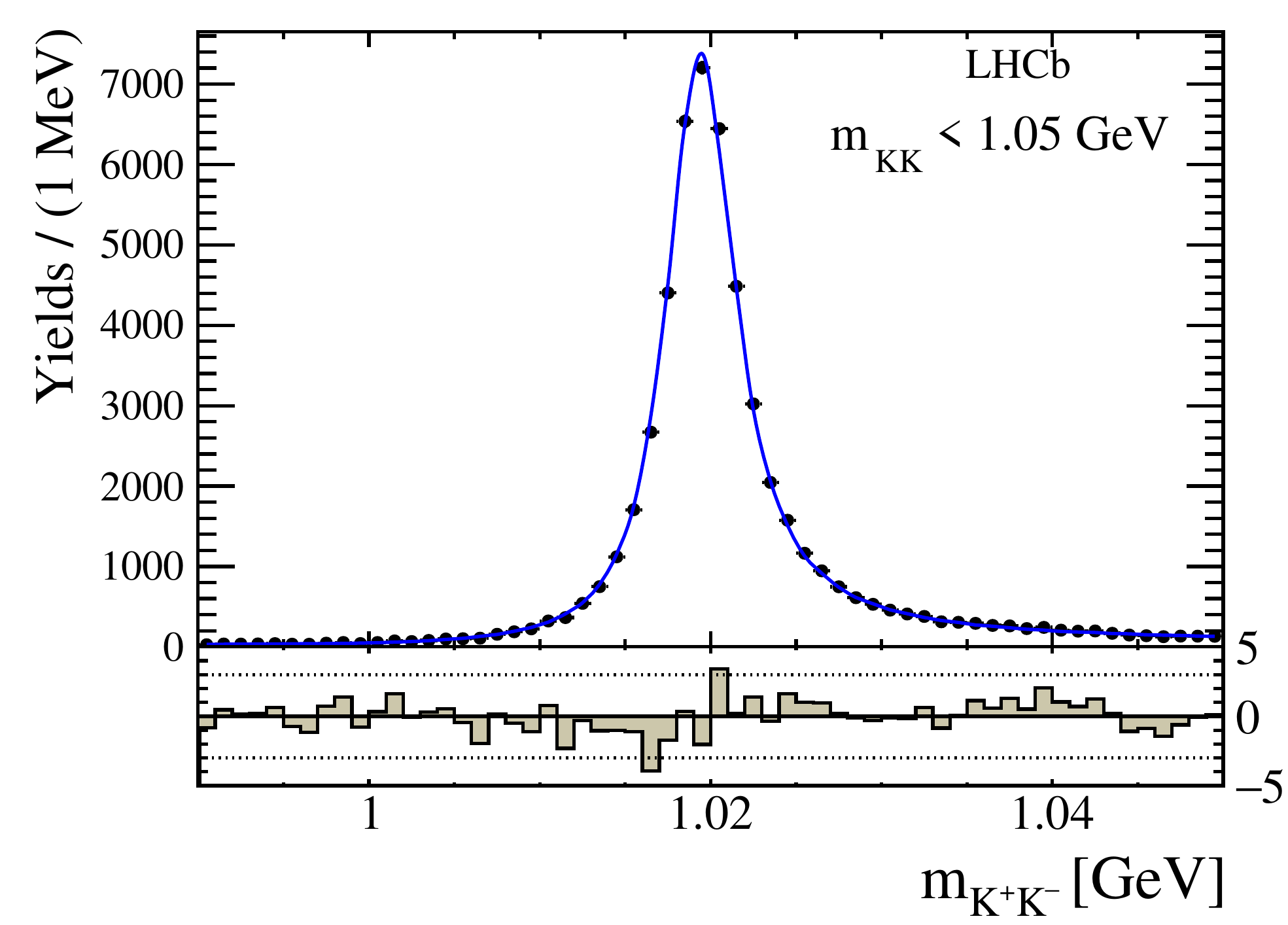}
\includegraphics[width=0.4\textwidth]{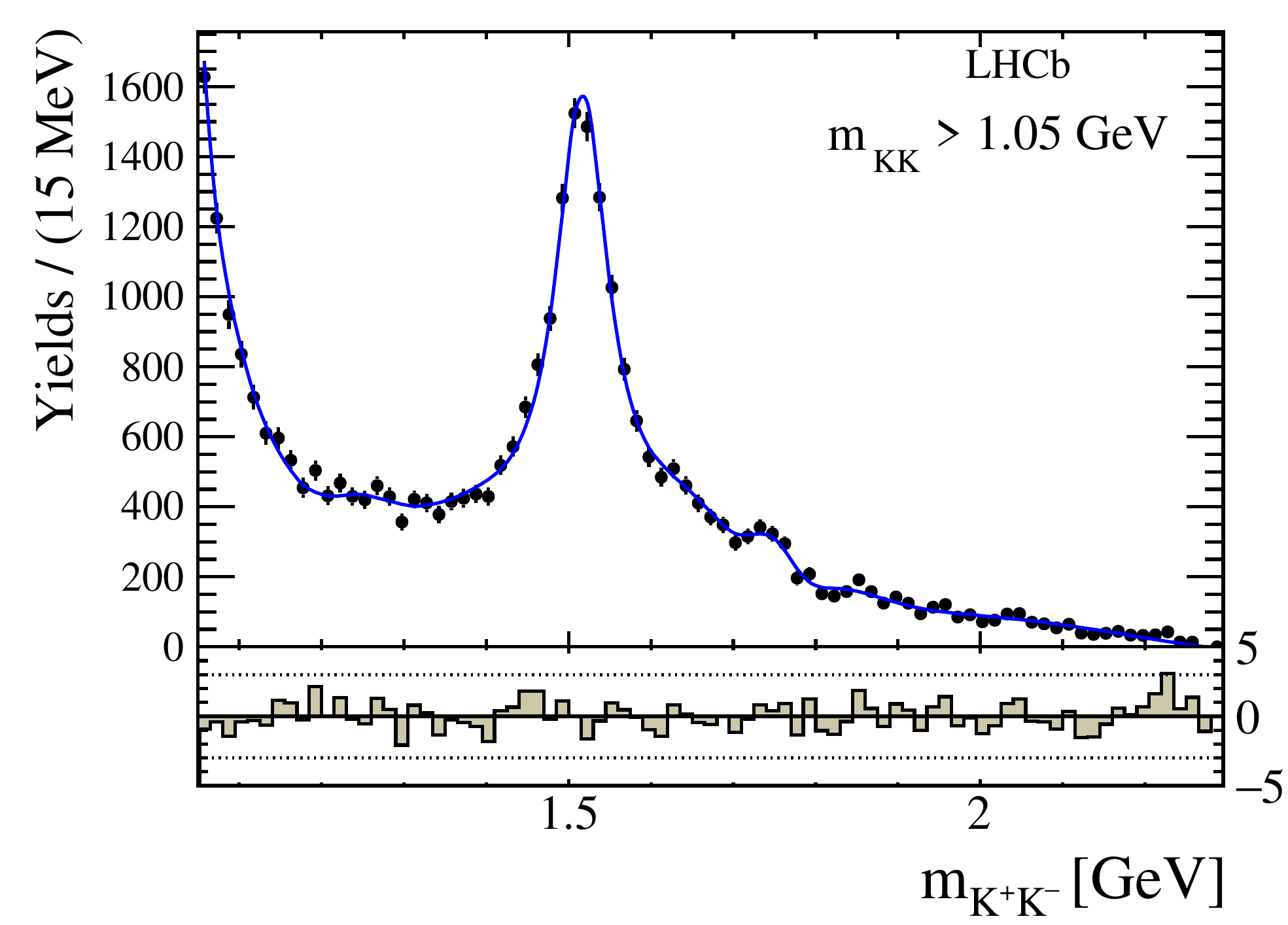}
\includegraphics[width=0.4\textwidth]{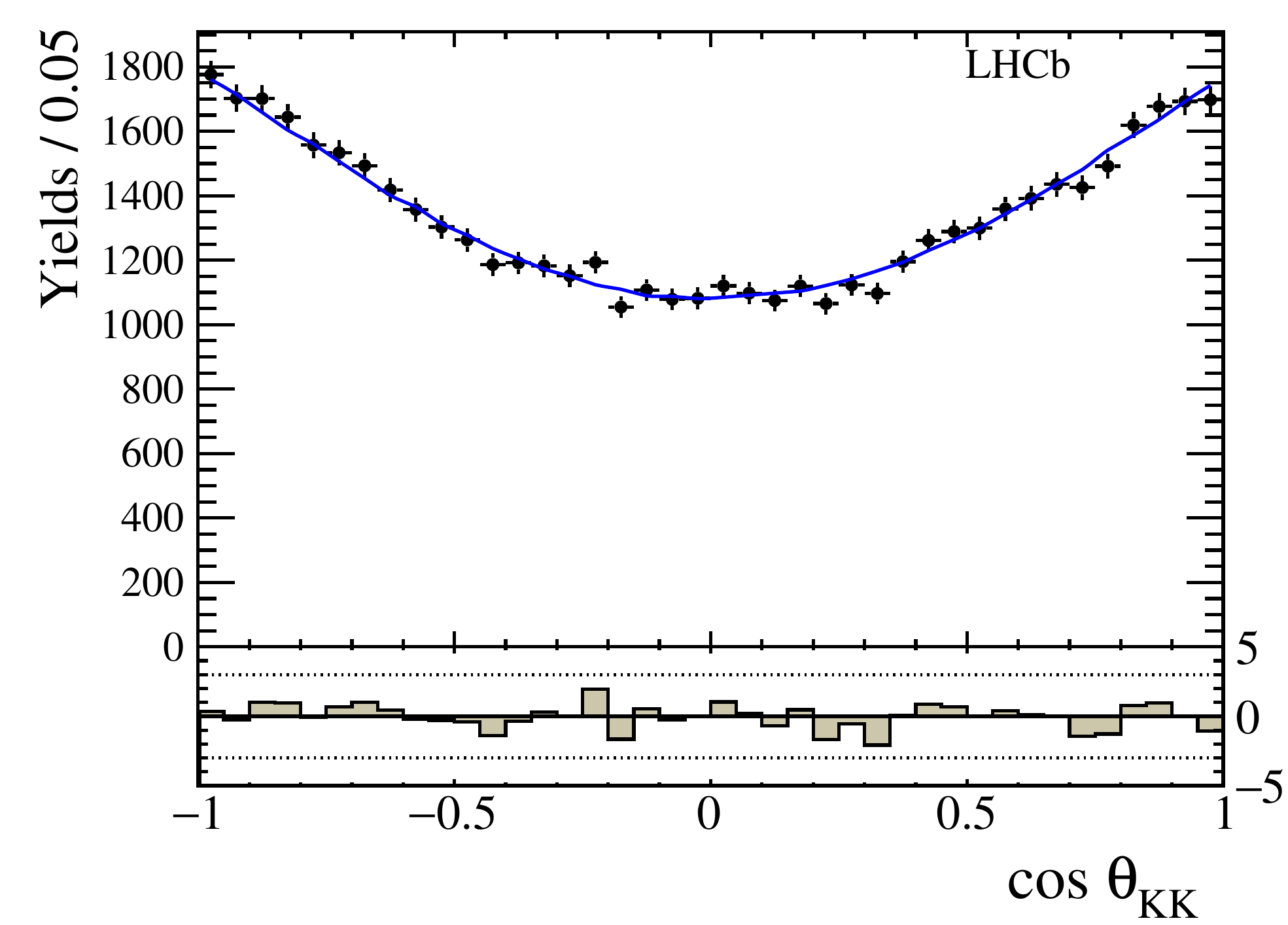}%
\includegraphics[width=0.4\textwidth]{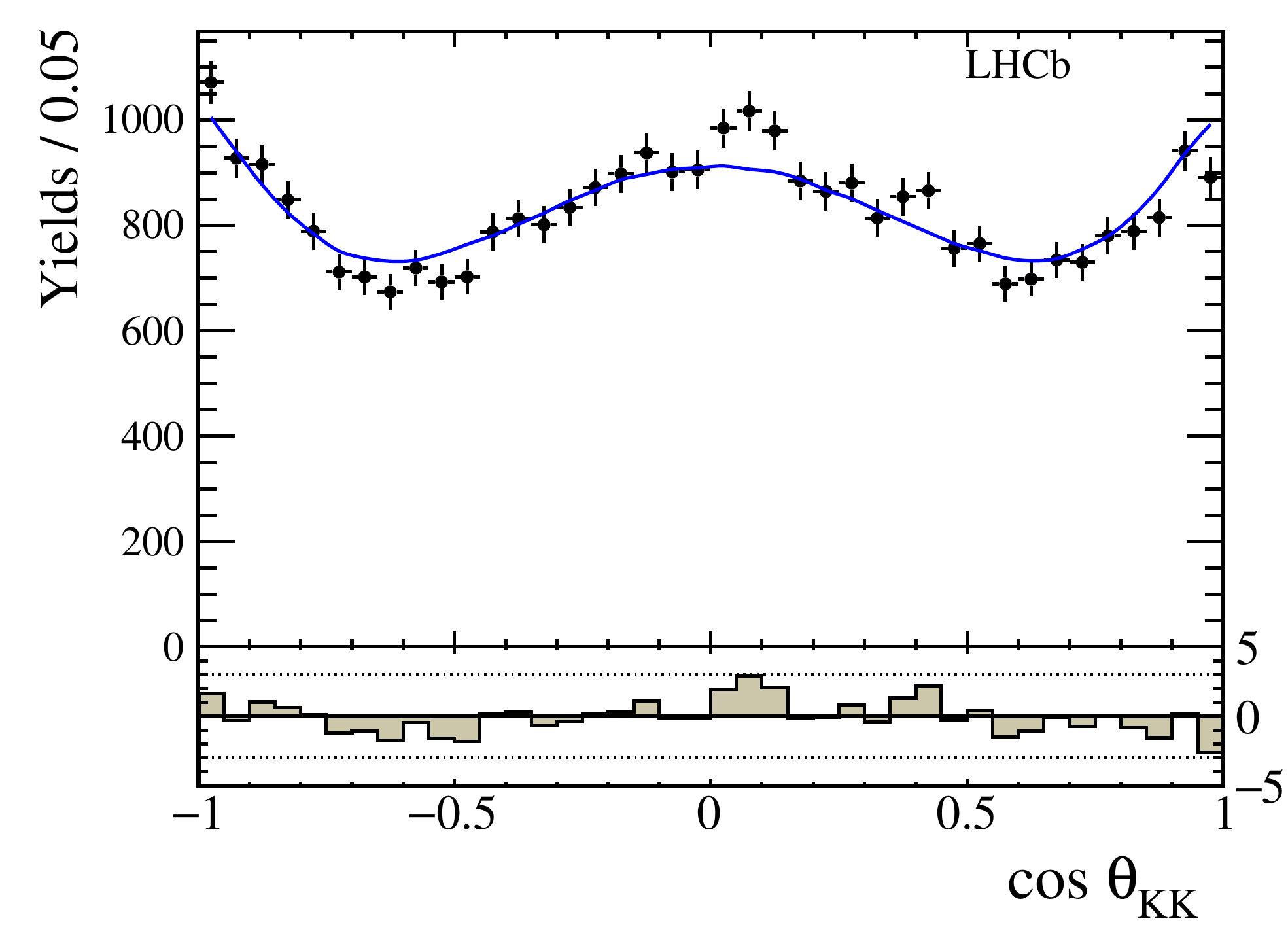}
\includegraphics[width=0.4\textwidth]{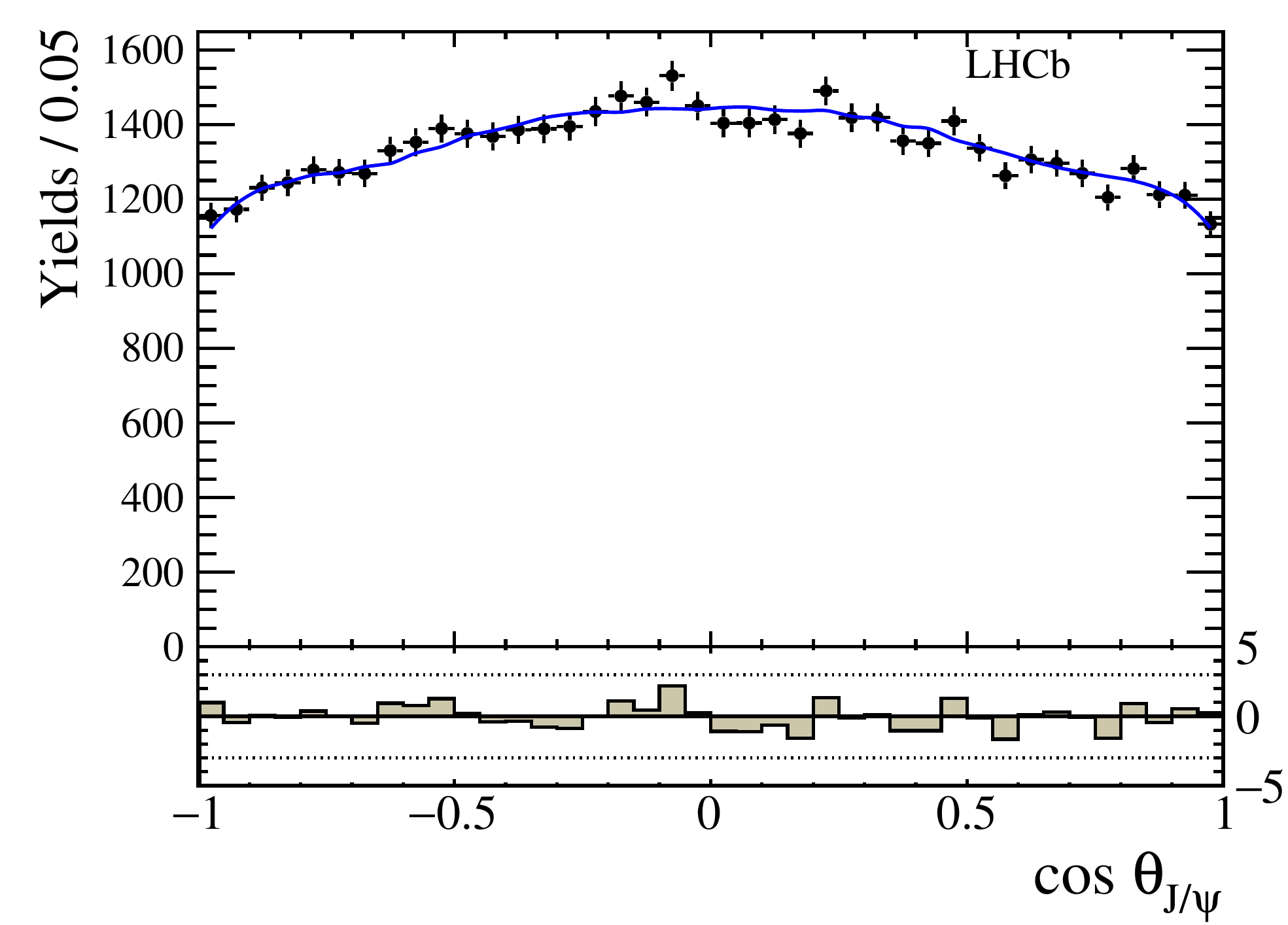}
\includegraphics[width=0.4\textwidth]{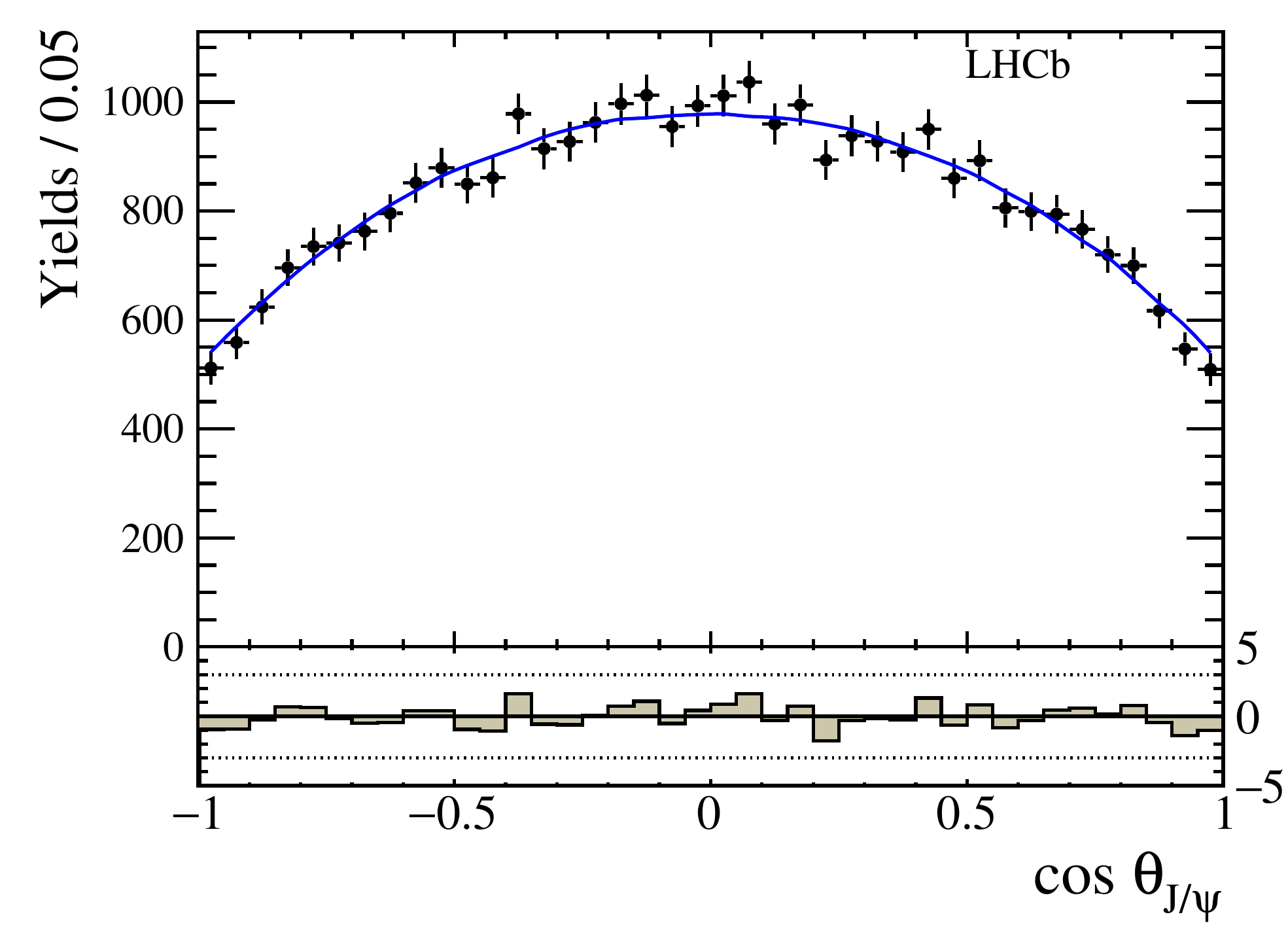}
\includegraphics[width=0.4\textwidth]{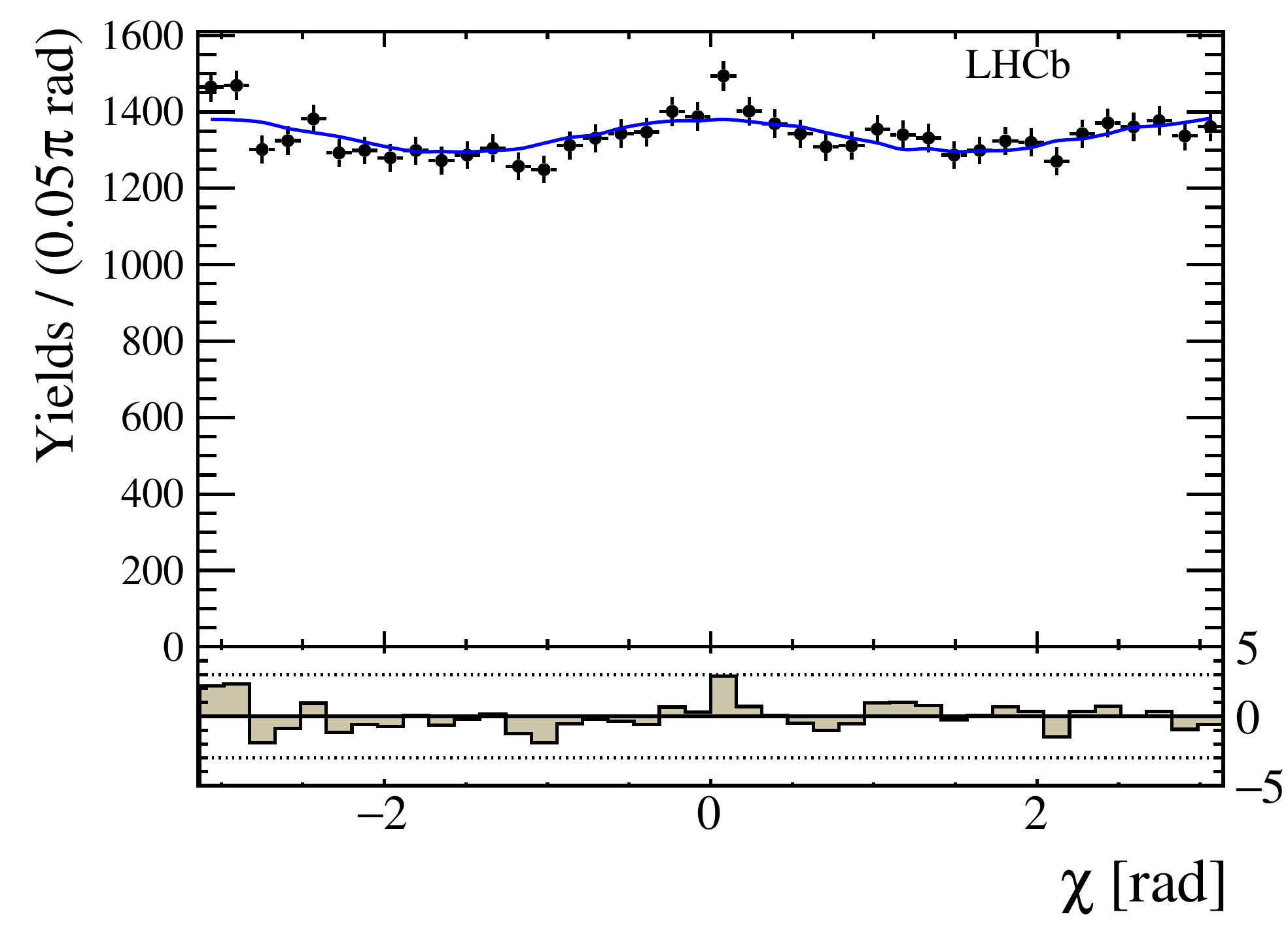}
\includegraphics[width=0.4\textwidth]{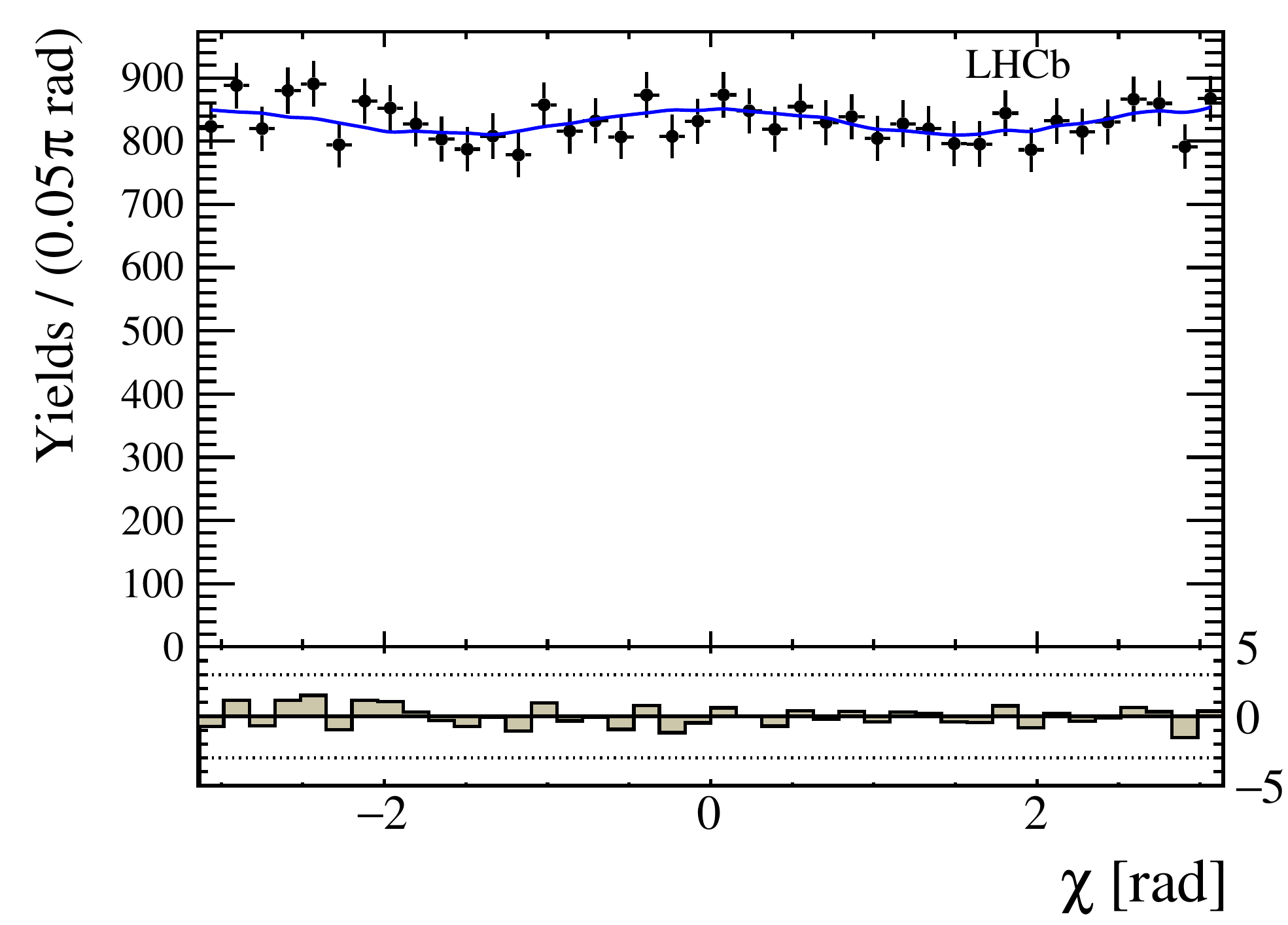}
\includegraphics[width=0.4\textwidth]{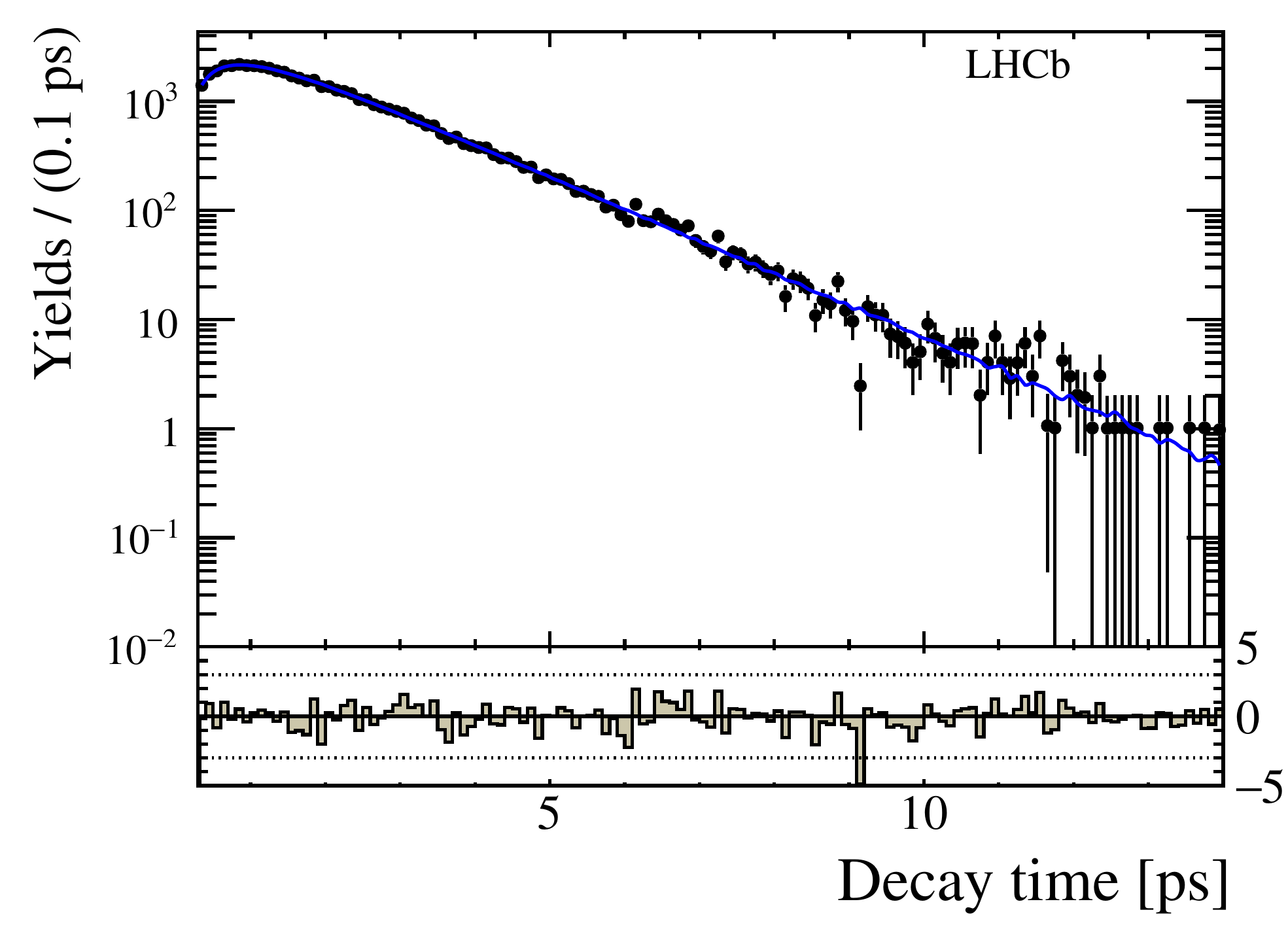}
\includegraphics[width=0.4\textwidth]{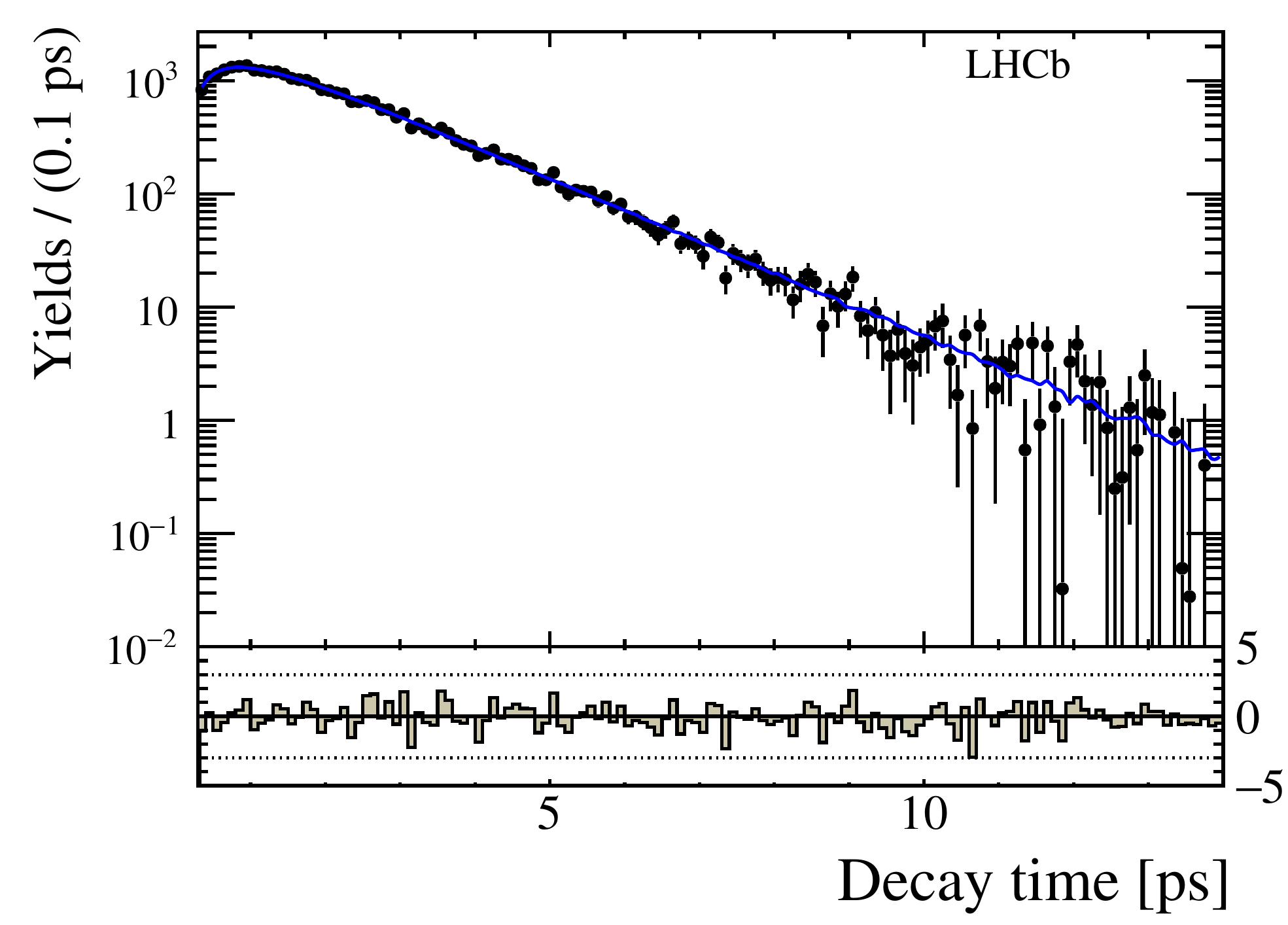}
\caption{\small Projections of the fitting variables in the (left) low-mass ($\phi(1020)$)  and (right) high-mass regions shown by the solid (blue) curves. The points with error bars are the data. At the bottom of each figure the differences between the data and the fit divided by the uncertainty in the data are shown. }\label{fig:fitphi}
\end{figure}

\begin{figure}[t]
\centering
\includegraphics[width=0.75\textwidth]{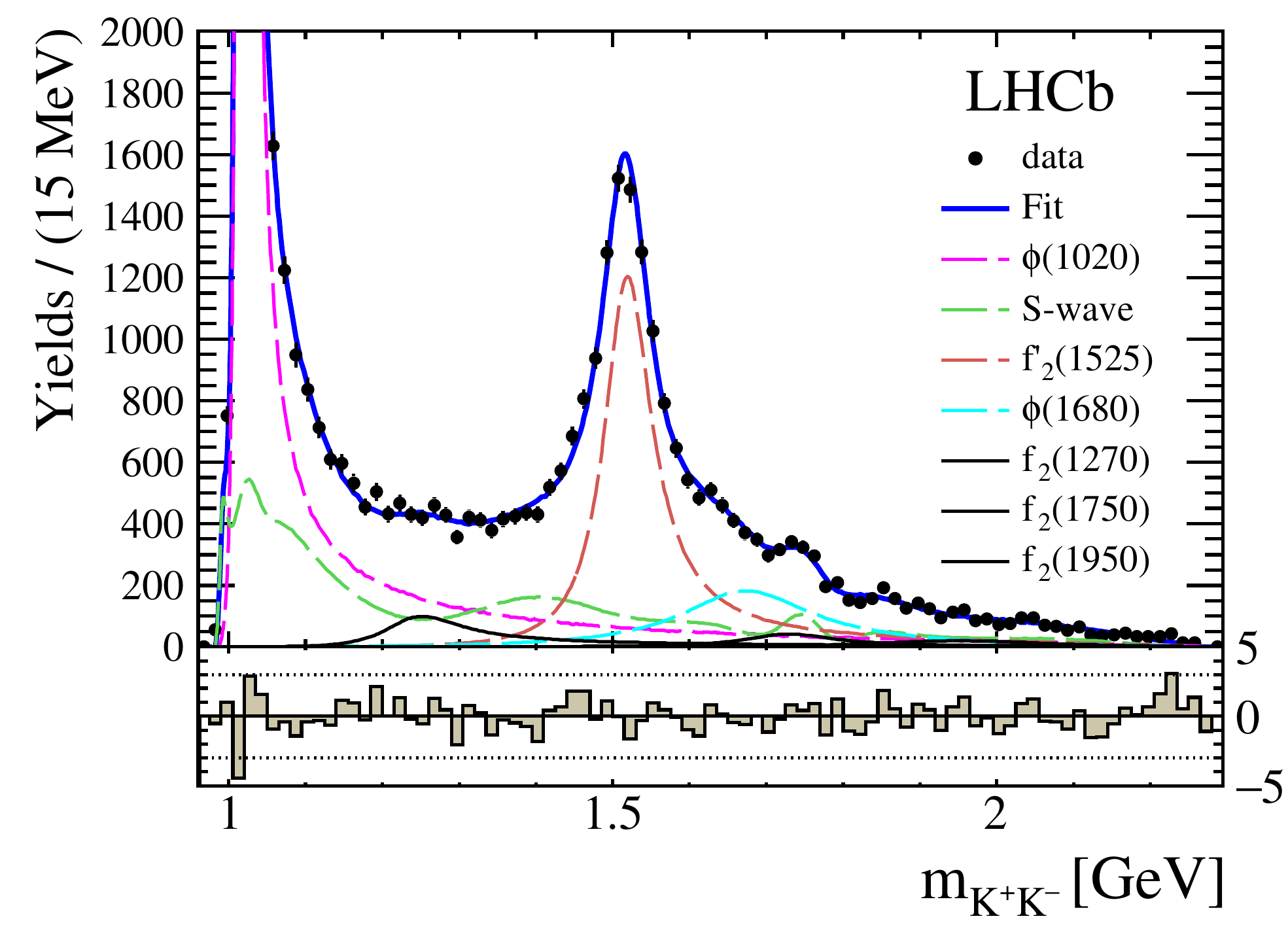}%
\caption{\small Fit projection of $\m$. The points represent the data; the resonances $\phi(1020)$, $f_2^\prime(1525)$, $\phi(1680)$ are shown by magenta, brown, and cyan long-dashed curves, respectively; the S-wave component is depicted by green long-dashed curves; the other $f_2$ resonances are described by black solid curves; and the total fit by a blue solid curve. At the bottom the differences between the data and the fit divided by the uncertainty in the data are shown. }
\label{Fig_7_new}
\end{figure}

The fit fractions are reported in Table~\ref{tab:fit2} and resonance phases in Table~\ref{tab:phfit}. Fit projections are shown in Fig.~\ref{fig:fitphi} for the $\phi(1020)$ region and above.  The fit reproduces the data in each of the projected variables. Each contributing component is shown in Fig.~\ref{Fig_7_new} as a function of $\m$.
To check the fit quality in the high $\m$ region, $\chi^2$ tests are performed. For $\m$ and  $\Omega$,  
$\chi^2$=1401 for 1125 bins (25 for $\m$, 5 for $\cos\angpi$, 3 for $\cos\angmu$ and 3 for $\chi$); for the two variables $\m$ and $\cos\angpi$, $\chi^2$=380 for 310 bins. The fit describes the data well. Note, adding the $f_2(1640)$ into the fit improves the $-2\ln{\cal L}$ by 0.4 with an additional 6 degrees of freedom, showing that this state is not observed.

\begin{table}[tb]
\centering
\caption{\small Fit results of the resonant structure.}\label{tab:fit2}
\def\arraystretch{1.2}
\begin{tabular}{lcccc}
\multirow{2}{*}{Component}& \multirow{2}{*}{Fit fraction (\%)}   & \multicolumn{3}{c}{Transversity fraction (\%)}\\
&  &
                    $0$ & $\parallel$& $\perp$\\\hline
$\phi(1020)$	&$	70.5	\pm	0.6\pm1.2	$&$	50.9	\pm	0.4	$&$	23.1	\pm	0.5	$&$	26.0	\pm	0.6	$\\				
$f_2(1270)$	&\,\,\,$	1.6	\pm	0.3	\pm0.2$&$	76.9	\pm	5.5	$&\,\,\,$	6.0	\pm	4.2	$&$	17.1	\pm	5.0	$\\				
$f_2^\prime(1525)$	&$	10.7	\pm	0.7	\pm0.9$&$	46.8	\pm	1.9	$&$	33.8	\pm	2.3	$&$	19.4	\pm	2.3	$\\			
$\phi(1680)$	&\,\,\,$	4.0	\pm	0.3 \pm0.3	$&$	44.0	\pm	3.9	$&$	32.7	\pm	3.6	$&$	23.3	\pm	3.6	$\\				
$f_2(1750)$	&$	0.59	\,_{-	0.16	}^{+	0.23	}\pm0.21$&\,\,$	58.2	\pm	13.9	$&\,\,\,$	31.7	\pm	12.4	$&$	10.1	\,_{-	6.1	}^{+	16.8	}$\\
$f_2(1950)$	&$	0.44	\,_{-	0.10	}^{+	0.15	}\pm0.14$&$	2.2	\,_{-	1.5	}^{+	6.7	}$&\,\,\,$	38.3	\pm	13.8	$&\,\,\,$	59.5	\pm	14.2	$\\
S-wave	&\,\,\,$	10.69	\pm	0.12	\pm0.57$&$	100			$&$	0			$&$	0			$\\							
 \end{tabular}
 \end{table}

 \begin{table}[b]
\centering
\caption{\small Fitted phase differences between two transversity states (statistical uncertainty only). Here the symbol $\phi$ refers to the components of the $\phi(1020)$ meson.}\label{tab:phfit}
\begin{tabular}{lrl}
States & \multicolumn{2}{c}{Phase difference ($^{\circ}$)} \\\hline
$	f_2(1270)^0	-	 \phi^\perp		$&$	139.5$\,&$\!\!\!\!\!	\pm \,	6.5	$\\
$	f_2^\prime(1525)^0	-	\phi^\perp		$&$	-167.9$\,&$\!\!\!\!\!	\pm\,	6.6	$\\
$	f_2(1750)^0	-	\phi^\perp		$&$	-251.5$\,&$\!\!\!\!\!	\pm	\,13.0	$\\
$	f_2(1950)^0	-	 \phi^\perp		$&$	-84.1$\,&$\!\!\!\!\!	\pm	\,42.1	$\\
$	\phi(1680)^0	-	 \phi^0		$&$	181.5$\,&$\!\!\!\!\!	\pm\,5.2	$\\
\hline
$	f_2(1270)^\perp	-	 \phi^0		$&$	100.5$\,&$\!\!\!\!\!	\pm\,	16.1	$\\
$	f_2^\prime(1525)^\perp	-	 \phi^0		$&$	-145.4	$\,&$\!\!\!\!\!\pm\,	9.2	$\\
$ f_2(1750)^\perp	-	 \phi^0		$&$	230.2	$\,&$\!\!\!\!\!\pm	\,36.1	$\\
$	 f_2(1950)^\perp	-	 \phi^0		$&$	116.7$\,&$\!\!\!\!\!	\pm	\,17.4	$\\
$	 \phi^\perp	-	 \phi^0		$&$	199.7$\,&$\!\!\!\!\!	\pm	\,7.6	$\\
$	 \phi(1680)^\perp	-	 \phi^\perp		$&$	134.0$\,&$\!\!\!\!\!	\pm	\,7.6	$\\
\hline									
									
$	 f_2(1270)^\parallel	-	 \phi^\perp		$&$	-140.3$\,&$\!\!\!\!\!	\pm	\,21.4	$\\
$	f_2^\prime(1525)^\parallel	-	 \phi^\perp		$&$	46.2	$\,&$\!\!\!\!\!\pm\,	7.9	$\\
$	 f_2(1750)^\parallel	-	 \phi^\perp		$&$	-27.5$\,&$\!\!\!\!\!	\pm	\,15.9	$\\
$	 f_2(1950)^\parallel	-	 \phi^\perp		$&$	3.8	$\,&$\!\!\!\!\!\pm	\,19.5	$\\
$	 \phi^\parallel	-	 \phi^0		$&$	195.4	$\,&$\!\!\!\!\!\pm	\,3.8	$\\
$	 \phi(1680)^\parallel	-	 \phi^0		$&$	-105.8	$\,&$\!\!\!\!\!\pm\,	8.9	$\\
\end{tabular}
\end{table}

\afterpage{\clearpage}

As a check a fit is performed allowing independent sets of \CP-violating parameters
$(|\lambda_i|,\phi_s^i)$: three sets for the three corresponding $\phi(1020)$ transversity
states, one for the $K^+K^-$ S-wave, one common to all three transversity
states of  the $f_2(1270)$, one for the  $f_2^\prime(1525)$, one for the $\phi(1680)$, and one for the
combination of the two high-mass $f_2(1750)$ and $f_2(1950)$ resonances.  In total, eight
sets of \CP-violating parameters are used instead of two sets in the nominal fit. The $-2\ln{\cal L}$ value is improved by 16 units with 12 additional parameters compared to the nominal fit, corresponding to the fact that all states have consistent \CP violation within 1.3\,$\sigma$. All values  of $|\lambda|$ are consistent with unity and $\phi_s$ differences of  the longitudinal $\phi(1020)$ component are consistent with zero, showing no dependence of \CP violation for the different states. 
 
\section{Systematic uncertainties}
\label{sec:systematic}
The systematic uncertainties are summarized for the physics parameters in Table~\ref{tab:sys1} and for the fit fractions in Table~\ref{tab:sys2}. They are small compared to the statistical ones for the \CP-violating parameters.
Generally, the largest contribution results from the resonance fit model.  The fit model uncertainties are determined by doubling the number of S-wave knots in the high $\m$ region, allowing the centrifugal barrier factors, of nominal value $1.5\gev^{-1}$ for $K^+K^-$ resonances and $5.0\gev^{-1}$ for the $\Bs$ meson \cite{LHCb-PAPER-2015-029}, to vary within 0.5--2 times of these values \cite{VonHippel:1972fg}. Additional systematic uncertainties are evaluated by increasing the orbital angular momentum between the \jpsi and the $K^+K^-$ system from the lowest allowed one, which is taken as the nominal value, and varying the masses and widths of contributing resonances by their uncertainties. The largest variation among those changes is assigned as the systematic uncertainty for resonance modelling. 
The effect of using the $m_0$  in the fit, rather than following the Belle approach using $\m$ is evaluated by redoing the fit. This change
worsens the $-2\ln{\cal L}$ by more than 100 units, which clearly shows the variation doesn't give a good fit; as a consequence, no systematic uncertainty is assessed. 
Differences resulting from the two conventions are comparable to the quoted modelling uncertainty for the \CP-violating parameters, but generally are larger than the quoted systematic uncertainties for the fit fractions of nonscalar resonances.


\begin{table}[t]
\centering
\caption{\small Absolute systematic uncertainties for the physics parameters determined from the high $\m$ region compared to the corresponding  statistical uncertainty. Here $M_0$ and $\Gamma_0$ refer to the uncertainties on the $f_2^\prime(1525)$ resonance mass and width.}\label{tab:sys1}
\begin{tabular}{lcccccr}
Source &	$\DGs$&	$\Gs$ & $|\lambda|$& $M_0$ &	$\Gamma_0$ &	$\phis~~\!$\\
$\times 10^{-3}$ &  [\invps]& [\invps] &  &  [\gev] & [\gev] & [rad] \\\hline
Resonance modelling	&$	6.9	$&$	1.9	$&$	5.5	$&$	1.1	$&$	3.6	$&$	23.6	$	\\
Efficiency	($\m$, $\Omega$)&$	3.0	$&$	0.9	$&$	0.5	$&$	0.1	$&$	0.7	$&$	3.4	$	\\	
Efficiency $t$&$	2.2	$&$	2.8	$&$	0.0	$&$	0.0	$&$	0.0	$&$	0.0	$	\\	
$\tau_\Bzb$	&$	1.4	$&$	2.0	$&$	0.0	$&$	0.0	$&$	0.0	$&$	0.0	$	\\	
$t$ resolution	&$	0.3	$&$	0.2	$&$	0.2	$&$	0.0	$&$	0.0	$&$	1.1	$	\\
Fit bias & 5.0  & 1.1 & - & - & - & -\,\,\,\\
$A_{\rm P}$	&$	0.1	$&$	0.3	$&$	1.4	$&$	0.0	$&$	0.0	$&$	4.0	$	\\
Tagging	&$	1.2	$&$	0.3	$&$	0.8	$&$	0.0	$&$	0.0	$&$	11.2	$\\
Background	&$	0.5	$&$	0.8	$&$	0.4	$&$	0.1	$&$	0.1	$&$	1.5	$	\\
$\sWeight{s}$	&$	1.1	$&$	0.1	$&$	0.5	$&$	0.1	$&$	0.4	$&$	21.4	$	\\
$\Bc$       & - &  $0.5$ & - &  - & - & - \,\,\,\\
\hline

Total syst.&$	9.6	$&$	4.3	$&$	5.7	$&$	1.1	$&$	3.7	$&$	34.2	$	\\

\hline
Stat.	&$	17.7	$&$	5.5	$&$	18.0	$&$	1.3	$&$	3.0	$&$	106.6	$	\\

\end{tabular}
\end{table}

\begin{table}[b]
\centering
\caption{\small Combined systematic and statistical uncertainties in the fit fractions using an absolute scale where the numbers are in units of \%. ``Res. modelling" refers to resonance modelling.}\label{tab:sys2}
\begin{tabular}{lccccccc}
Source & $\phi(1020)$& S-wave & $f_2^\prime(1525)$&$\phi(1680)$& $f_2(1270)$& $f_2(1750)$& $f_2(1950)$\\\hline
Res. modelling&	$	0.99	$&$	0.57	$&$	0.73	$&$	0.27	$&$	0.21	$&$	0.21	$&$	0.13	$\\
Efficiency	&$	0.58	$&$	0.06	$&$	0.48	$&$	0.12	$&$	0.04	$&$	0.03	$&$	0.01	$\\
Background	& $	0.06	$&$	0.01	$&$	0.06	$&$	0.02	$&$	0.02	$&$	0.01	$&$	0.00	$\\	
$\sWeight{s}$	&$	0.11	$&$	0.02	$&$	0.16	$&$	0.05	$&$	0.02	$&$	0.05	$&$	0.04	$\\				 			
\hline															
Total syst.&	$	1.15	$&$	0.57	$&$	0.89	$&$	0.30	$&$	0.21	$&$	0.21	$&$	0.14	$\\\hline
\rule{0pt}{1em} Statistical & $0.62$ & $0.12$& $0.67$ &$0.32$ &$0.27$ &$_{-0.16}^{+0.23}$ &$_{-0.10}^{+0.15}$ \\
\end{tabular}
\end{table}

The sources of uncertainty for the modelling of the efficiency variation of the three angles and $\m$ include the statistical uncertainty from simulation, and the efficiency correction due to the differences in kinematic distributions between data and simulation for $\Bs$ decays. The former is estimated by repeating the fit to the data 100 times. In each fit, the efficiency parameters are resampled according to the corresponding covariance matrix determined from simulation. For the latter, the efficiency used by the nominal fit is obtained by weighting the distributions of $p$ and $\pt$ of the kaon pair and $\Bs$ meson to match the data. Such weighting is removed to assign the corresponding systematic uncertainty. 

The uncertainties due to the $\Bz$ lifetime and decay time efficiency determination are estimated.  Each source is evaluated by adding to the nominal fit an external correlated multidimensional Gaussian constraint, either given by the fit to the $\Bzb \to \jpsi \Kstarzb$ sample with varying $\tau_{\Bd}=1.520\pm0.004$\ps~\cite{PDG2016}, or given by the fit to simulation for the decay time efficiency correction, \ie ${\varepsilon_{\rm sim}^{\Bs}(t)}/{\varepsilon_{\rm sim}^{\Bd}(t)}$ in Eq.~(\ref{eqn:acc}). A systematic uncertainty is given by the difference in quadrature of the statistical uncertainties for each physics parameter between the nominal fit and the alternative fit with each of these constraints. The uncertainties due to the decay time acceptance are found to be negligible for the fit fraction results. 

The sample of prompt $\jpsi$ mesons combined with two kaon candidates is used to calibrate the per-candidate decay-time error. This method is validated by simulation. Since the detached selection, pointing angle and BDTG requirements cannot be applied to the calibration sample,  the simulations show that the calibration overestimates the resolution for $\Bs$ decays after final selection by about 4.5\%.  Therefore, a 5\% variation of the widths, and the uncertainty of the mean value are used to estimate uncertainty of the time resolution modelling. The average angular resolution is 6\,mrad for all three decay angles. This is small enough to have only negligible effects on the analysis.

A large number of pseudoexperiments is used to validate the fitter and check potential biases in the fit outputs. Biases on $\Gs$ and $\DGs$, $20\%$ of their statistical uncertainties, are taken as systematic uncertainties. Calibration parameters of the flavour-tagging algorithm and the $\Bs$--$\Bsb$ production asymmetry $A_{\rm P}=(1.09\pm2.69)\%$~\cite{LHCb-PAPER-2014-042} are fixed. The systematic uncertainties due to the calibration of the tagging parameters or the value of $A_{\rm P}$ are given by the difference in quadrature between the statistical uncertainty for each physics parameter between the nominal fit and an alternative fit where the tagging parameters or $A_{\rm P}$ are Gaussian-constrained by the corresponding uncertainties. 
Background sources are tested by varying the decay-time acceptance of the injected reflection backgrounds,  changing these background yields by 5\%, and also varying the $\Lb$ lifetime.

To evaluate the uncertainty of the \sPlot method that requires the fit observables being uncorrelated with the variable $m(\jpsi K^+K^-)$ used to obtain the \sWeight{s}, two variations are performed to obtain new \sWeight{s}, and the fit is repeated. The first consists of changing the number of $|\cos\angmu|$ bins.  In the nominal fit, the \sWeight{s} are determined by separate fits in four $|\cos\angmu|$ bins for the event candidates, as significant variations of signal invariant mass resolution are seen as a function of the variable. In another variation of the analysis  starting with the nominal number of $|\cos\angmu|$ bins the decay time dependence is explored, since the combinatorial background may have a possible variation as a function of $m(\jpsi K^+K^-)$. Here  the decay time is further divided into three intervals. The larger change on the physics parameter of interest is taken as a systematic uncertainty. 

About 0.8\% of the signal sample is expected from the decays of $\Bc$ mesons \cite{LHCb-PAPER-2013-044}. Neglecting the $\Bc$ contribution in the nominal fit leads to a negligible bias of 0.0005\invps for $\Gs$~\cite{LHCb-PAPER-2014-059}. The correlation matrix with both statistical and systematic uncertainties is shown in Table~\ref{tab:cormartixforHigh}.

\begin{table}[h]
\centering
\caption{The correlation matrix from the high-mass region fit, taking into account both statistical and systematic uncertainties.}
\label{tab:cormartixforHigh}
\vspace{-0.2cm}
\begin{tabular}{ccccc}
& $\Gamma_s$ &$\Delta\Gamma_s$ &$\phi_s$ &$|\lambda|$ \\\hline 
$\Gamma_s$& $+1.00$&$+0.54$&$+0.02$&$-0.03$\\
$\Delta \Gamma_s$&&$+1.00$&$+0.04$&$-0.06$\\
$\phi_s$&&&$+1.00$&$-0.14$\\
$|\lambda|$&&&&$+1.00$\\
\end{tabular}
\end{table}

\section{Conclusions}
\label{sec:conclusion}
\begin{table}[b]
\centering
\caption{The correlation matrix taking into account both statistical and systematic uncertainties for the combination of the three measurements $\Bs\to\jpsi K^+K^-$ for $\m>1.05$~GeV,  $\m<1.05$~GeV, and $\jpsi\pi^+\pi^-$.}
\label{tab:cormartixforcomb}
\vspace{-0.2cm}
\begin{tabular}{ccccc}
& $\Gamma_s$ &$\Delta\Gamma_s$ &$\phi_s$ &$|\lambda|$ \\\hline 
$\Gamma_s$& $+1.00$&$-0.13$&$-0.01$&$\phantom{+}0.00$\\
$\Delta \Gamma_s$&&$+1.00$&$-0.05$&$\phantom{+}0.00$\\
$\phi_s$&&&$+1.00$&$-0.04$\\
$|\lambda|$&&&&$+1.00$\\
\end{tabular}
\end{table}
We have studied $\Bs$ and $\Bsb$  decays into the $\jpsi \Kp \Km$ final state using a time-dependent amplitude analysis. In the $\m>1.05$\gev region we determine
\begin{align*}
\phi_s &= 119\pm107\pm34 {\rm \, mrad,}\\
|\lambda| &= 0.994\pm0.018\pm0.006,\\
\Gs &= 0.650\pm0.006\pm0.004 \invps,\\
\DGs&=0.066\pm0.018\pm0.010 \invps.
\end{align*}

Many resonances and a S-wave structure have been found.  Besides the $\phi(1020)$ meson these include the $f_2(1270)$, the $f_2'(1525)$, the $\phi(1680)$, the $f_2(1750)$, and the $f_2(1950)$ mesons.  The presence of the $f_2(1640)$ resonance is not confirmed. The measured \CP-violating parameters of the individual resonances are consistent. 
The $f_2^\prime(1525)$ mass and width are determined as $1522.2\pm1.3\pm1.1$\mev and $78.0\pm3.0\pm3.7$\mev, respectively. The fit fractions of the resonances in $\Bs\to\jpsi \Kp\Km$ are also determined, and shown in Table~\ref{tab:fit2}.  These results supersede our previous measurements~\cite{LHCb-PAPER-2012-040}. 

The combination with the previous results from $\Bs$ decays in the $\phi(1020)$ region~\cite{LHCb-PAPER-2014-059} gives
\begin{align*}
\phi_s &= -25\pm45\pm8 {\rm \, mrad,}\\
|\lambda| &= 0.978\pm0.013\pm0.003,\\
\Gs &= 0.6588\pm0.0022\pm0.0015 \invps,\\
\DGs&=0.0813\pm0.0073\pm0.0036 \invps.
\end{align*}
The two results are consistent within $1.1\sigma$. A further combination is performed by including the $\phis$ and $|\lambda|$ measurements from $\Bs$ and $\Bsb$ decays into $\jpsi \pip\pim$~\cite{LHCb-PAPER-2014-019}, which results in $\phis = 1\pm37$\,mrad and $|\lambda|=0.973\pm0.013$, where $\Gs$ and $\DGs$ are unchanged. The correlation matrix is shown in Table~\ref{tab:cormartixforcomb}. The measurement of the \CP-violating phase $\phis$ is in agreement with the SM prediction $-36.5_{-1.2}^{+1.3}$\,mrad~\cite{Charles:2015gya}. These new combined results supersede our combination reported in Ref.~\cite{LHCb-PAPER-2014-059}.

\newpage

\section*{Acknowledgements}

\noindent We express our gratitude to our colleagues in the CERN
accelerator departments for the excellent performance of the LHC. We
thank the technical and administrative staff at the LHCb
institutes. We acknowledge support from CERN and from the national
agencies: CAPES, CNPq, FAPERJ and FINEP (Brazil); MOST and NSFC (China);
CNRS/IN2P3 (France); BMBF, DFG and MPG (Germany); INFN (Italy);
NWO (The Netherlands); MNiSW and NCN (Poland); MEN/IFA (Romania);
MinES and FASO (Russia); MinECo (Spain); SNSF and SER (Switzerland);
NASU (Ukraine); STFC (United Kingdom); NSF (USA).
We acknowledge the computing resources that are provided by CERN, IN2P3 (France), KIT and DESY (Germany), INFN (Italy), SURF (The Netherlands), PIC (Spain), GridPP (United Kingdom), RRCKI and Yandex LLC (Russia), CSCS (Switzerland), IFIN-HH (Romania), CBPF (Brazil), PL-GRID (Poland) and OSC (USA). We are indebted to the communities behind the multiple open
source software packages on which we depend.
Individual groups or members have received support from AvH Foundation (Germany),
EPLANET, Marie Sk\l{}odowska-Curie Actions and ERC (European Union),
Conseil G\'{e}n\'{e}ral de Haute-Savoie, Labex ENIGMASS and OCEVU,
R\'{e}gion Auvergne (France), RFBR and Yandex LLC (Russia), GVA, XuntaGal and GENCAT (Spain), Herchel Smith Fund, The Royal Society, Royal Commission for the Exhibition of 1851 and the Leverhulme Trust (United Kingdom).

\clearpage

{\noindent\normalfont\bfseries\Large Appendix}

\appendix

\label{sec:Supplementary-App}

\section{Angular moments}
\label{sec:moments}

\def \t {\theta_{KK}}
\def \A {{\cal A}}

\begin{figure}[!hbtp]
\centering
\includegraphics[width=0.33\textwidth]{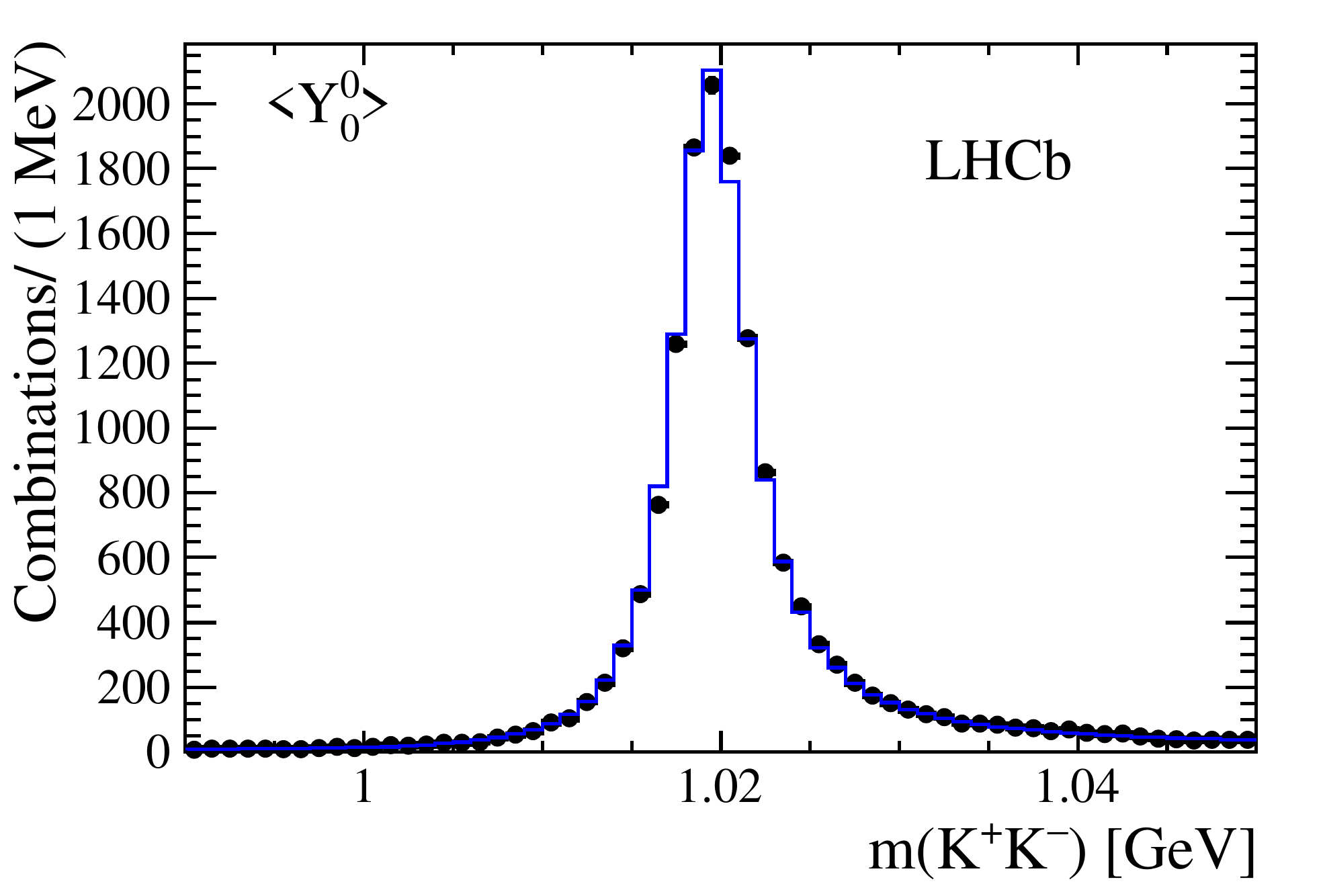}%
\includegraphics[width=0.33\textwidth]{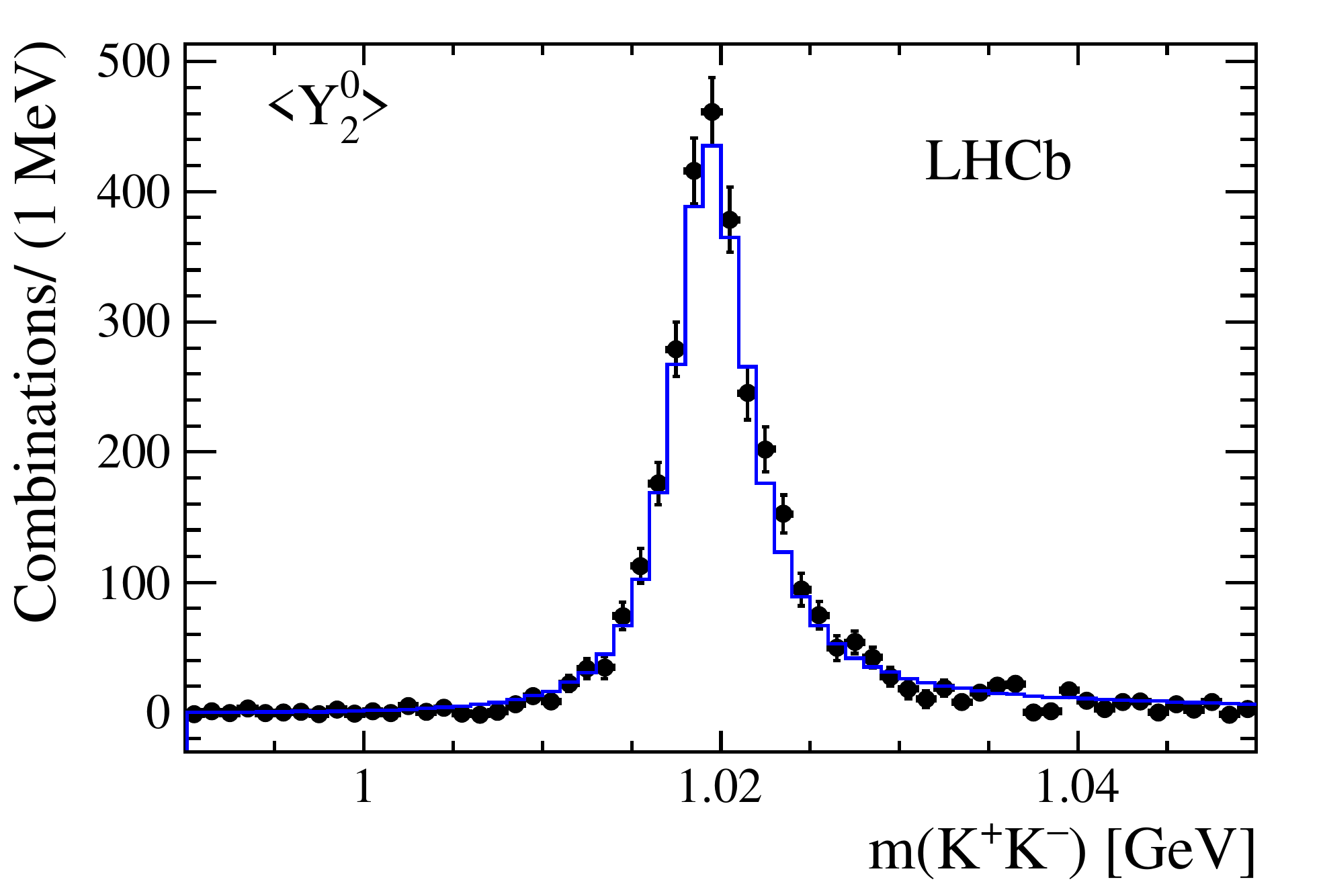}%
\includegraphics[width=0.33\textwidth]{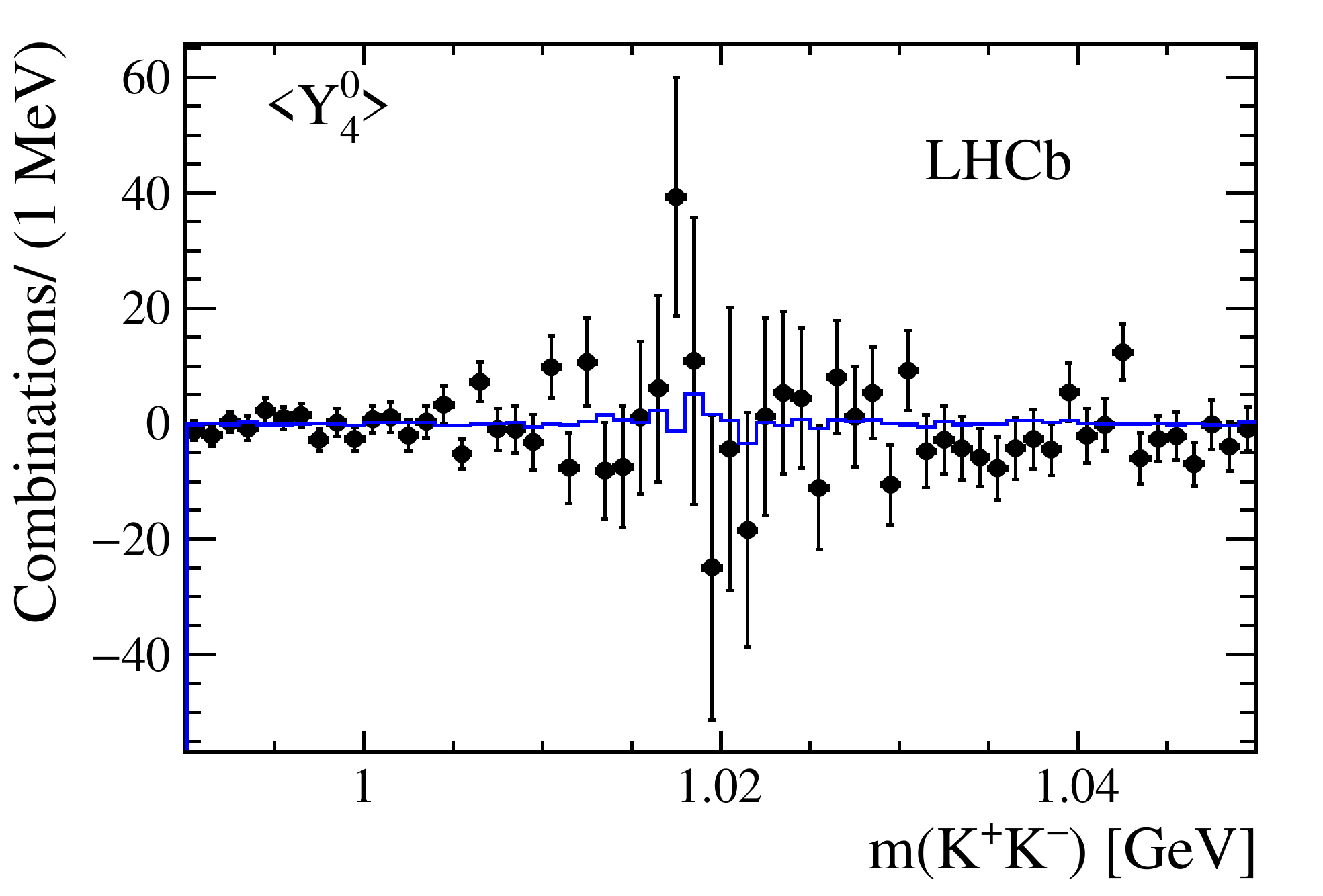}
\caption{The $K^+K^-$ mass dependence of the spherical harmonic moments of $\cos \theta_{KK}$ 
in the region of the $\phi(1020)$ resonance after efficiency corrections and background subtraction. The points with error bars are the data points and the (blue) lines are derived from the fit model.   }
\label{fig:Sph_phi}
\end{figure}

\begin{figure}[!hbtp]
\centering
\includegraphics[width=0.33\textwidth]{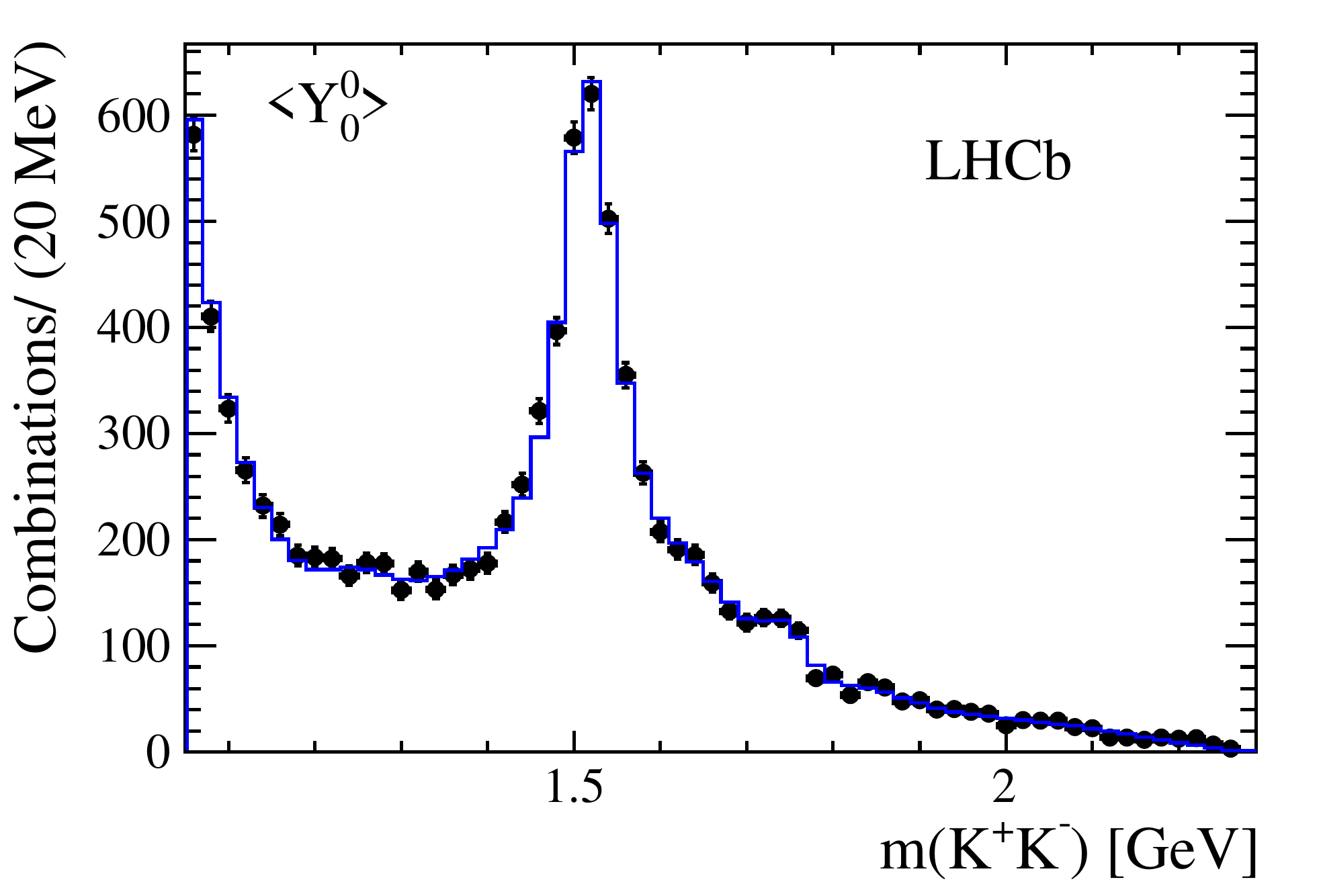}%
\includegraphics[width=0.33\textwidth]{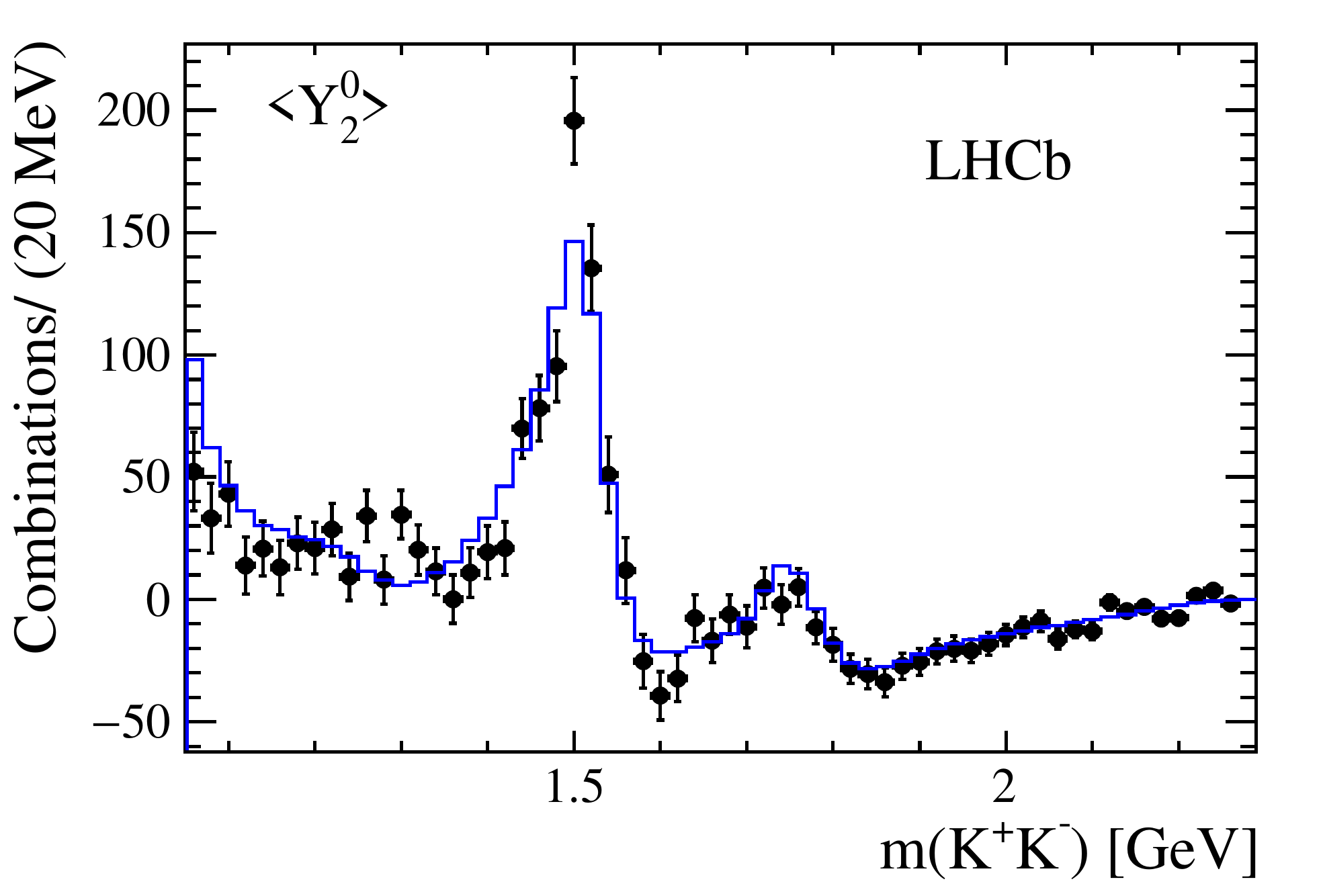}%
\includegraphics[width=0.33\textwidth]{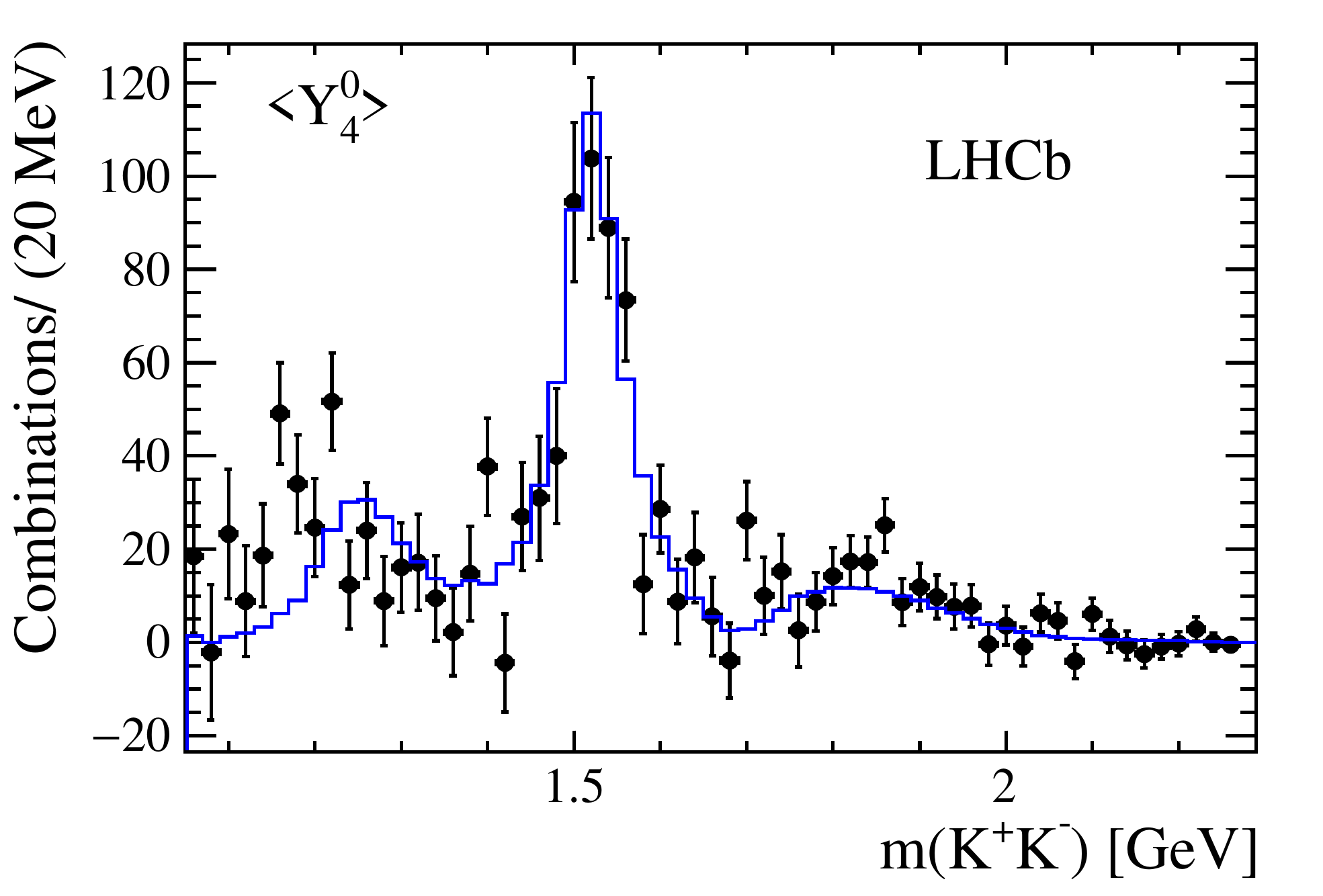}
\caption{The $K^+K^-$ mass dependence of the spherical harmonic moments of $\cos \theta_{KK}$ above the $\phi(1020)$ resonance region after efficiency corrections and background subtraction.  The points with error bars are the data points and the (blue) lines are derived from the fit model.}
\label{fig:Sph}
\end{figure}

We define the moments 
 $\langle Y_{\ell}^0\rangle$, as the efficiency-corrected and
background-subtracted $\Kp\Km$ invariant mass distributions, weighted by the $\ell$th spherical harmonic functions of the cosine of the helicity angle $\t$. The
moment distributions provide an additional way of visualizing the presence of different resonances and
their interferences, similar to a partial wave analysis. Figures \ref{fig:Sph_phi} and \ref{fig:Sph} show the distributions of the even angular moments for the events around $\pm 30$ MeV of $\phi(1020)$ mass peak and those above the $\phi(1020)$, respectively. The general interpretation of the even moments is that  $\langle Y^0_0\rangle$ is the efficiency-corrected and background-subtracted event distribution,  $\langle Y^0_2\rangle$  reflects the sum of P-wave, D-wave and the interference of S-wave and D-wave amplitudes,  and $\langle Y^0_4\rangle$  the D-wave. The average of $\Bs$ and $\Bsb$ decays cancels the interference terms that involve P-wave amplitudes. This causes the odd moments to sum to zero. 

The fit results reproduce the moment distributions relatively well. For the region near the $\phi(1020)$, the $p$-values are 3\%, 3\%, 48\% for the $\ell$=0, 2, 4 moments, respectively. For the high mass region, the $p$-values are 37\%, 0.2\% 0.5\% for the $\ell$=0, 2, 4 moments, respectively.


\clearpage


\newpage
\addcontentsline{toc}{section}{References}
\setboolean{inbibliography}{true}
\bibliographystyle{LHCb}
\bibliography{main}



 
\newpage
\centerline{\large\bf LHCb collaboration}
\begin{flushleft}
\small
R.~Aaij$^{40}$,
B.~Adeva$^{39}$,
M.~Adinolfi$^{48}$,
Z.~Ajaltouni$^{5}$,
S.~Akar$^{59}$,
J.~Albrecht$^{10}$,
F.~Alessio$^{40}$,
M.~Alexander$^{53}$,
S.~Ali$^{43}$,
G.~Alkhazov$^{31}$
% ...

{\footnotesize \it
% ...
$ ^{5}$Clermont Universit{\'e}, Universit{\'e} Blaise Pascal, CNRS/IN2P3, LPC, Clermont-Ferrand, France\\
% ...
$ ^{10}$Fakult{\"a}t Physik, Technische Universit{\"a}t Dortmund, Dortmund, Germany\\
% ...
$ ^{31}$Petersburg Nuclear Physics Institute (PNPI), Gatchina, Russia\\
% ...
$ ^{39}$Universidad de Santiago de Compostela, Santiago de Compostela, Spain\\
$ ^{40}$European Organization for Nuclear Research (CERN), Geneva, Switzerland\\
% ...
$ ^{43}$Nikhef National Institute for Subatomic Physics, Amsterdam, The Netherlands\\
% ...
$ ^{48}$H.H. Wills Physics Laboratory, University of Bristol, Bristol, United Kingdom\\
% ...
$ ^{53}$School of Physics and Astronomy, University of Glasgow, Glasgow, United Kingdom\\
% ...
$ ^{59}$University of Cincinnati, Cincinnati, OH, United States\\
% ...
}
\end{flushleft}

%
%

\end{document}